\documentclass[]{aa}
\usepackage{natbib}
\usepackage[utf8]{inputenc}
\usepackage{graphicx}
\usepackage{subcaption}
\usepackage{txfonts}
\usepackage{color}
\usepackage{amsmath}
\usepackage{comment}

\usepackage{scalerel}
\usepackage{tikz}
\usetikzlibrary{svg.path}

\definecolor{orcidlogocol}{HTML}{A6CE39}
\tikzset{
  orcidlogo/.pic={
    \fill[orcidlogocol] svg{M256,128c0,70.7-57.3,128-128,128C57.3,256,0,198.7,0,128C0,57.3,57.3,0,128,0C198.7,0,256,57.3,256,128z};
    \fill[white] svg{M86.3,186.2H70.9V79.1h15.4v48.4V186.2z}
                 svg{M108.9,79.1h41.6c39.6,0,57,28.3,57,53.6c0,27.5-21.5,53.6-56.8,53.6h-41.8V79.1z M124.3,172.4h24.5c34.9,0,42.9-26.5,42.9-39.7c0-21.5-13.7-39.7-43.7-39.7h-23.7V172.4z}
                 svg{M88.7,56.8c0,5.5-4.5,10.1-10.1,10.1c-5.6,0-10.1-4.6-10.1-10.1c0-5.6,4.5-10.1,10.1-10.1C84.2,46.7,88.7,51.3,88.7,56.8z};
  }
}
\newcommand\orcidicon[1]{\href{https://orcid.org/#1}{\mbox{\scalerel*{
\begin{tikzpicture}[yscale=-1,transform shape]
\pic{orcidlogo};
\end{tikzpicture}
}{|}}}}

\usepackage[hidelinks]{hyperref} %<--- Load after everything else

\definecolor{gaolvBlue}{RGB}{0,134,228}
\definecolor{gaolvOrange}{RGB}{235,98,0}
\definecolor{gaolvGray}{gray}{0.8}

\newlength{\subfiglen}

\newcommand{\unit}[1]{\,\mathrm{#1}}
\newcommand{\tensor}[1]{\mathsf{#1}}
\bibpunct{(}{)}{;}{a}{}{,} % to follow the A&A style

\begin{document}
	\title{Modeling disk fragmentation and multiplicity in massive star formation}
	\titlerunning{Modeling disk fragmentation and multiplicity in massive star formation}
	\author{G.~André~Oliva %\inst{1}
	\and
	Rolf~Kuiper %\inst{1}
	}

	\institute{Institute for Astronomy and Astrophysics, University of Tübingen, Auf der Morgenstelle 10, 72076, Tübingen, Germany\\
		\email{andree.oliva@uni-tuebingen.de} \orcidicon{0000-0003-0124-1861}  \\
		\email{rolf.kuiper@uni-tuebingen.de}  \orcidicon{0000-0003-2309-8963}
	}

	% ORCiD Rolf: 0000-0003-2309-8963
	% ORCiD André: 0000-0003-0124-1861

	\abstract{
	There is growing evidence that massive stars grow by disk accretion similar to their low-mass counterparts.
	Early in evolution, these disks can achieve masses which are comparable to the current stellar mass, and hence, the forming disks are highly susceptible for gravitational fragmentation.
	}{
	We investigate the formation and early evolution of an accretion disk around a forming massive protostar, focussing on its fragmentation physics.
	For this, we follow the collapse of a molecular cloud of gas and dust, the formation of a massive protostar, the formation of its circumstellar disk, and the formation and evolution of the disk fragments.
	%, i.e.~self-gravity and radiative cooling.
	}{
	We use a grid-based self-gravity-radiation-hydrodynamics code including a sub-grid module for stellar evolution and dust evolution.
	On purpose, we do not use a sub-grid module for fragmentation such as sink particles to allow for all paths of fragment formation and destruction, but instead keeping the spatial grid resolution high enough to properly resolve the physical length scales of the problem, namely the pressure scale height and Jeans length of the disk.
	Simulations are performed on a grid in spherical coordinates with a logarithmic spacing of the grid cells in the radial direction and a cosine-distribution of the grid cells in the polar direction, focusing the spatial resolution on the disk midplane. Because of that, roughly 25\% of the total number of grid cells, corresponding to $\sim$ 26 million grid cells, are used to model the disk physics. They constitute the highest resolution simulations performed up to now on disk fragmentation around a forming massive star with the physics considered here.
	For a better understanding of the effects of spatial resolution and to compare our high-resolution results with previous lower resolution studies in the literature, we perform the same simulation for five different resolutions, each of them with a factor of two lower resolution than the predecessor run.
	}{
	The cloud collapses and a massive (proto)star is formed in its center, surrounded by a fragmenting Keplerian-like accretion disk with spiral arms. The fragments have masses of $\sim 1 \unit{M_\odot}$, and their continuous interactions with the disk, spiral arms and other fragments results in eccentric orbits. Fragments form hydrostatic cores, surrounded by secondary disks with spiral arms that also produce new fragments. We identified several mechanisms of fragment formation, interaction and destruction. Central temperatures of the fragments can reach the hydrogen dissociation limit, form second Larson cores and evolve into companion stars. Based on this, we study the multiplicity predicted by the simulations and find $\sim 6$ companions at different distances from the primary: from possible spectroscopic multiples, to companions at distances between $1000$ and $2000\unit{au}$.
	}
	{}


	\maketitle

	% - - - - - - - - - - - - - - - - - - - - - - - - - - - - - - - - - - - - - - - -
\section{Introduction}

During the formation of massive stars ($\gtrsim 8\unit{M_\odot}$), radiation pressure becomes important against the gravity of the collapsing molecular cloud. The formation of an accretion disk with polar outflows provides a mechanism for circumventing the radiation pressure barrier \citep[see, e.g.,][]{2010ApJ...722.1556K} and allow the forming star to become massive. This disk is expected to fragment and produce companion stars.

There is growing observational evidence that supports this scenario. Observations of disks around massive (proto-)stars are reported by, for example, \cite{2015ApJ...813L..19J},  \cite{2016MNRAS.462.4386I}, \cite{2017A&A...602A..59C}, \cite{2018ApJ...860..119G} and \cite{2019A&A...627L...6M}. Some of these disks have also been shown to be Keplerian-like.

Moreover, there is evidence that, early in evolution, these disks gain enough mass to become self-gravitating, form spiral arms and fragment. \cite{2018ApJ...869L..24I} observed a fragmented Keplerian disk around the proto-O star G11.92-0.61 MM1a, with a fragment MM1b in the outskirts of the disk, at $\sim 2000\unit{au}$ from the primary. \cite{2017A&A...603A..10B} reported a smaller disk-like structure around the central object in the G351.77-0.54 high-mass hot core, and a fragment at about $\sim 1000\unit{au}$. \cite{2020A&A...634L..11J} have also observed spiral arms and instability in a disk of radius $\sim 1000\unit{au}$ around the O-type star AFGL 4176 mm1. \cite{2019A&A...627L...6M} reports substructures in a Keplerian disk around the O-type (proto-)star G17.64+0.16.

Several studies with three-dimensional hydrodynamical simulations, including radiation transport and self gravity, of a collapsing cloud aiming to form a massive star, have been performed. Some of them lead to a massive star surrounded by a fragmenting accretion disk, and some form stars via filament accretion (we offer a literature review, including methods, initial conditions and results in Sect. \ref{S: literature}). \cite{2009Sci...323..754K}, \cite{2010ApJ...708.1585K}, \cite{2016MNRAS.463.2553R}, \cite{2016ApJ...823...28K}, \cite{2012MNRAS.420..613G} and others, used adaptive mesh refinement (AMR) grids and sub-grid sink particle models. In the case of \cite{2012MNRAS.420..613G}, as many as $\sim 400$ sink particles were reported in some of their runs, or in the case of \cite{2016MNRAS.463.2553R}, up to $\sim 30$; in contrast, \cite{2016ApJ...823...28K} report none. These big disparities in the number of possible formed companions under similar conditions provokes questions on the role that spatial resolution and the sink particle algorithms used in these studies play on the final outcome of the system. In Sections \ref{S: numerical} and \ref{S: literature}, we explore this matter in detail, and find that higher resolution is needed to resolve the Jeans length than used in previous studies.

Accretion in fragmented disks is expected to not be a smooth process, but characterized by episodic accretion of some fragments. The release of gravitational potential energy creates an increase in luminosity, i.e., an accretion burst. Accretion bursts offer an explanation to the luminosity burst events in regions of massive star formation which have been reported in, e.g., \cite{2017ApJ...837L..29H}, \cite{2017NatPh..13..276C} and \cite{2019ATel12446....1S}. Recently, \cite{2020NatAs.tmp..144C} reported on the observation of disk substructures associated with an accretion burst event, thus providing a link between the two phenomena. The simulations in \cite{2017MNRAS.464L..90M, 2018MNRAS.473.3615M, 2019MNRAS.482.5459M} show accretion bursts; they performed an extensive analysis on the accretion process, intensity and frequency of the bursts.

Massive stars do not typically form in isolation. Studies on multiplicity in more evolved systems of massive stars show that a large portion of them are spectroscopic binaries \citep[see, e.g.,][]{2012Sci...337..444S, 2014ApJS..213...34K, 2015A&A...580A..93D}. The observations of fragmenting accretion disks suggest the possibility that close binaries may form by disk fragmentation and later inward migration of the companions \citep{2018MNRAS.473.3615M}.

In this study, we present the highest-resolution self-gravity-radiation-hydrodynamical simulations up to date on disk fragmentation around a forming massive (proto)star. We continue the approach taken by \cite{2018MNRAS.473.3615M}, in which no sink particles are set on purpose, so that fragmentation is described self-consistently as the interplay of self-gravity of the gas, cooling of the gas, shear of the disk and other gravito-radiation-hydrodynamical processes. Instead, a time-independent grid whose spatial resolution scales logarithmically with radius was used, allowing for very high resolution in the areas of most interest. We set up a series of simulations with increasing resolutions, and explore the physical processes that occur in fragmentation (how fragments evolve and interact), the properties of the fragments (orbits, mass and temperatures), their fate (whether they evolve further and eventually become companion stars), and the effects that resolution has on fragment statistics.

The outcome of our simulation was already used in \cite{2019A&A...632A..50A} for producing synthetic observations with ALMA and NOEMA. In that paper, many observationally-relevant quantities were computed, including 1.37 mm continuum images, integrated intensity and intensity-weighted peak velocity maps of $\mathrm{CH_3 CN(12_4-11_4)}$, position--velocity plots, and Toomre $Q$ maps, taking into account several inclinations. We refer the reader to that reference for an ample discussion on the observability of the results presented here.

This paper is organized as follows. Sections \ref{S: simulations} and \ref{S: pp-method} present the methodology of both the simulations, and the post-processing of the generated data. Sections \ref{S: overview} and \ref{S: accretion disk}--\ref{S: companion} present our results and a discussion of disk fragmentation, the fragments themselves and companion formation. Section \ref{S: numerical} contains a convergence study of our results, and Sect. \ref{S: literature}, a literature review and comparison of our results with previous studies.

	% - - - - - - - - - - - - - - - - - - - - - - - - - - - - - - - - - - - - - - - -
\section{Setup of the simulations} \label{S: simulations}

\subsection{Physics}
	In order to model the problem, we consider the three-dimensional collapse of an initially axially-symmetric molecular cloud of total mass $200 \unit{M_\odot}$ and radius $0.1\unit{pc}$. The initial density and rotational profiles are given in Sect. \ref{S: initial boundary cond}.

	The cloud dynamics are modeled as an ideal gas with the hydrodynamics equations:

	\begin{align}
		&\frac{\partial \rho}{\partial t} + \vec{\nabla} \cdot (\rho\vec{v}) = 0 \\
		&\frac{\partial}{\partial t} (\rho \vec{v})  +  \vec{\nabla} \cdot  (\rho \vec v\otimes \vec v + P\tensor{I}) = \rho\vec a_\text{ext}\\
		&\frac{\partial E}{\partial t} +\vec \nabla \cdot  \left( (E+P) \vec v \right) = \rho \vec v \cdot \vec a_\text{ext}
 	\end{align}

	\noindent where $\rho, P, E$ are the density, pressure and energy density, respectively; $\vec v$ is the velocity, and $\vec a_\text{ext}$ is the acceleration source term. These equations are solved with the hydrodynamics module of the numerical grid code Pluto \citep{2007ApJS..170..228M}, with additional modules for handling radiation transport and self-gravity.

	The gravity of the forming massive star, the self-gravity of the fluid and the radiation forces are incorporated into the code via the acceleration source term, such that $\vec a_\text{ext} = \vec a_\star + \vec a_\text{sg} + \vec a_\text{rad} $, and
	\begin{align*}
	 	\vec a_\star &= -\frac{GM_\star}{r^2} \vec e_r \\
	 	\vec a_\text{sg} &= -\vec \nabla \Phi_\text{sg}\text{, where} \quad	\vec \nabla^2 \Phi_\text{sg} = 4\pi G \rho \\
		\vec a_\text{rad} &= -\frac{\vec \nabla \cdot \vec F_\star}{\rho c} \vec e_r - \kappa_R \frac{D \vec \nabla E_R}{c} \\
	\end{align*}

% 2010 Fast and freq dep (39)

	The variables involved in $\vec a_\text{rad}$ are detailed below. The Poisson equation of self gravity is solved by means of a diffusion ansatz. More details on the implementation of the Poisson solver for the self-gravity module are given in \cite{2010ApJ...722.1556K}.

	The gas is assumed to be calorically perfect, that is, governed by the calorical equation of state  $P = (\gamma - 1) E_\text{int}$ (where $\gamma = 5/3$, the value for $\text{H}_\text{2}$ at low temperatures, and $ E_\text{int}$ is the internal energy density).

	Radiation transport is incorporated by the method described in \cite{2010A&A...511A..81K}, but advanced to a two-temperature approach, which we will summarize in the rest of this section. We treat the star as an emitter of radiation; the dust and gas present in the cloud absorb it, and re-emit it, in addition to the thermal radiation due purely to hydrodynamic compression. This treatment allows us to divide the total radiation flux  $\vec F_\text{tot}$ into two contributions: the flux from the star, $\vec F_\star$, and the flux from thermal (re-)emission from dust and gas, $\vec F$.

	Stellar irradiation is solved by means of the time-independent radiation transport equation (with the valid assumption that the photon travel time is small compared to the time step of the  simulation, and only considering absorption), which yields

	\begin{equation}\label{E: rad-transp-star}
	\vec F_\star (\nu, r) = \vec F_\star(\nu, R_\star) \left(\frac{R_\star}{r}\right)^2  e^{-\tau(\nu,r)}, \text{where } \tau(r,\nu) =  \int_{R_\star}^r \kappa(\nu,r) \rho   dr
	\end{equation}
	is the optical depth; $\kappa(\nu,r)$, the frequency-dependent opacity; and $R_\star$, the stellar radius. This equation is integrated for each ray direction and frequency bin in each time step of the hydrodynamical simulation.

	The frequency-dependent opacity $ \kappa(\nu) $ is the sum of the dust and gas opacities. For the dust opacities, we use the tabulated values by \cite{1993ApJ...402..441L}, including 79 frequency bins, and we calculate the local evaporation temperature of the dust grains by using the formula of \cite{2005A&A...438..899I}. The gas opacity is set to a constant value of $0.01 \unit{cm^2\, g^{-1}}$. We also require the flux of the forming star at the stellar radius $R_\star$, for which we use tabulated evolutionary tracks for accreting high-mass stars, calculated by \cite{2009ApJ...691..823H}.

	For the remaining thermal (re-)emission contribution, $\vec F$ is assumed to be frequency-independent. First, we take the zeroth moment of the radiation transport equation

	\begin{equation}\label{E: rad-transp-dust}
		 \frac{\partial E_\text{rad}}{\partial t} - \vec \nabla \cdot \vec F =  \rho \kappa_P (4\pi B_\text{rad} - c E_\text{rad})
	\end{equation}

	where $B_\text{rad} = aT^4$ is the black-body energy density, $c$ is the speed of light in vacuum and $\kappa_P$ is the Planck mean opacity. This equation is solved by means of the flux-limited diffusion (FLD) approximation, which consists on  setting $ \vec F = -D \vec \nabla E_\text{rad} $, that is, assuming that radiation transport can be treated as a diffusion problem. The diffusion constant is $ D = \lambda c/\rho \kappa_R $, where $\lambda$ is the flux limiter, and $\kappa_R$ is the Rosseland mean opacity. The change of internal energy of the system is

	\begin{equation} \label{E: eint}
	\frac{\partial E_\text{int}}{\partial t} = 	-  \rho \kappa_P (4\pi B_\text{rad} - c E_\text{rad}) - \vec \nabla \cdot \vec F_\star
	\end{equation}

	where, for the ideal gas, $ E_\text{int} = c_v \rho T $. We solve equations \eqref{E: rad-transp-dust} and \eqref{E: eint} for the unknowns $ E_\text{rad} $ and $ T $ by using the so called two-temperature linearization approach, described in \cite{2011A&A...529A..35C}.

\subsection{Geometry}
	\begin{table}
		\centering
		\begin{tabular}{c c c c c c c}
				run & $n_r$ & $n_\theta$ & $n_\phi$  & $n_\text{disk}$ & $\Delta x_{30}$ & $\Delta x_{1500}$ \\ \hline \hline
				x1 & 34 & 11 & 32 & $7.0\cdot 10^3$ & 6.36 & 359 \\
				x2 & 67 & 21 & 64 & $5.3 \cdot 10^4$ & 3.07 & 167 \\
				x4 & 134 & 41 & 128 & $4.2\cdot 10^5$ &1.50 & 77.8 \\
				x8 & 268 & 81 & 256 & $3.3\cdot 10^6$ &  0.74 & 37.5 \\
				x16 & 536 & 161 & 512 & $2.6\cdot 10^7$ & 0.368 & 18.4
		\end{tabular}
		\caption{Designation code of each simulation (column 1), number of cells for each coordinate (columns 2--4); approximate number of cells in the region $r<1500\unit{au}$, i.e., the approximate number of cells that resolve the disk's physics (column 5); cell size at $r=30\unit{au}$ (column 6), cell size at $r=1500\unit{au}$ (column 7). $\Delta x_{30}$ is also the minimum cell size of the computational domain in the disk's midplane, and $\Delta x_{1500}$ represents roughly the cell size in the outer region of the disk.}
		\label{T: resolutions}
	\end{table}

	The equations described in the previous section are solved in a three-dimensional, spherical, time-independent grid, in 5 different resolutions that are detailed in Table~\ref{T: resolutions}. The computational domain extends from $r=30\unit{au}$ to $r=20\,626.5 \unit{au}$ ($=0.1\unit{pc}$). The coordinate grid is built as follows: the radial coordinate increases logarithmically from the inner boundary; the polar angle varies with the cosine function, so that maximum resolution is achieved in the midplane ($z=0$); and the azimuth is uniformly discretized. In the midplane the cells are approximately cubical, with two example cell sizes given as $\Delta x$ in Table~\ref{T: resolutions}.

	This choice of coordinates is explained and justified in more detail in \cite{2011ApJ...732...20K}, but in a nutshell, a spherical grid guarantees strict angular momentum conservation around the central massive protostar, and the logarithmic grid in $r$ allows for a focus on the phenomena closer to the forming massive star, while saving computational power. An explanation on how well does this choice of grids resolve the phenomena studied here is presented in Sect. \ref{S: resolution}, as well as comparison to the grid choices of previous studies.

\subsection{Initial and boundary conditions} \label{S: initial boundary cond}

	The initial density profile is spherically symmetric, and has the general form
	\[ \rho(r,t=0) = \rho_0 \left(\frac{r}{r_0}\right)^{\beta_\rho} \]
	We choose $\beta_\rho=-3/2$, and from the total mass and radius of the cloud, $\rho_0$ is determined to be $\approx 4.79\cdot 10^{-12} \unit{g\,cm^{-3}}$ at $r_0=1\unit{au}$. The value of $\beta_\rho$ was chosen based on the results of previous simulations and observations of massive dense cores (see the discussion in Sect. \ref{S: literature-addphysics}).

	The initial angular velocity is given by the profile
	\[ \Omega(R, t=0) = \Omega_0 \left( \frac{R}{R_0}\right)^{\beta_\Omega}\] 
	where $R$ is the cylindrical radius, and we choose $\beta_\Omega = -3/4$. The ratio of kinetic energy to gravitational potential energy is set to $5\%$, which fixes the normalization parameter $\Omega_0$ to $\approx 9.84\cdot 10^{-11} \unit{\,s^{-1}} $ at $R_0 = 10\unit{au}$. The normalization process is described in more detail in  \cite{2018MNRAS.473.3615M}, as well as a parameter scan for other choices of $\beta_\Omega$. The selection of $\beta_\Omega = \beta_\rho/2$ keeps the ratio of kinetic energy to gravitational energy independent of the radius of the cloud; the value for this ratio in turn was chosen in accordance to the typical values found in, e.g., \cite{1993ApJ...406..528G} and \cite{2014ApJ...785...42P}. The initial radial and polar velocities are set to zero.

	In summary, we give a low initial angular momentum to the cloud, so it forms a disk while collapsing. In a more realistic situation, inhomogeneities in the density profile may be present, as well as some initial sub-sonic turbulence and magnetic fields. The aim of this work, however, is to study fragmentation of the disk due to the development of Toomre instabilities in the accretion disk that forms in the process, and hence our choice of initial conditions. A more detailed discussion on the effects of the initial conditions and the inclusion of additional physical effects on the outcome of the simulations can be found in Sect. \ref{S: literature-addphysics}.

	 Both the inner and outer boundaries of the computational domain are semi-permeable, that is, matter is allowed to leave the computational domain but it can not re-enter it. For the purposes of the calculations involving the forming massive star, all the mass in the sink cell is considered as accreted by the central massive protostar forming in it.

	 The initial temperature is $10\unit{K}$, uniform across the computational domain, and the initial dust-to-gas mass ratio (used in the calculation of $ \kappa(\nu) $) is chosen to be 1\% \citep{2007ApJ...663..866D}.

		% - - - - - - - - - - - - - - - - - - - - - - - - - - - - - - - - - - - - - - - -
\section{Overview of the temporal evolution of the system} \label{S: overview}

		In the next sections, we use the data from runs x8 and x16 as a reference for the description and analysis of our results. A discussion of the results for the other runs can be found in Sect. \ref{S: numerical}.

	\begin{figure*}
	\centering
	\includegraphics[width=0.95\textwidth]{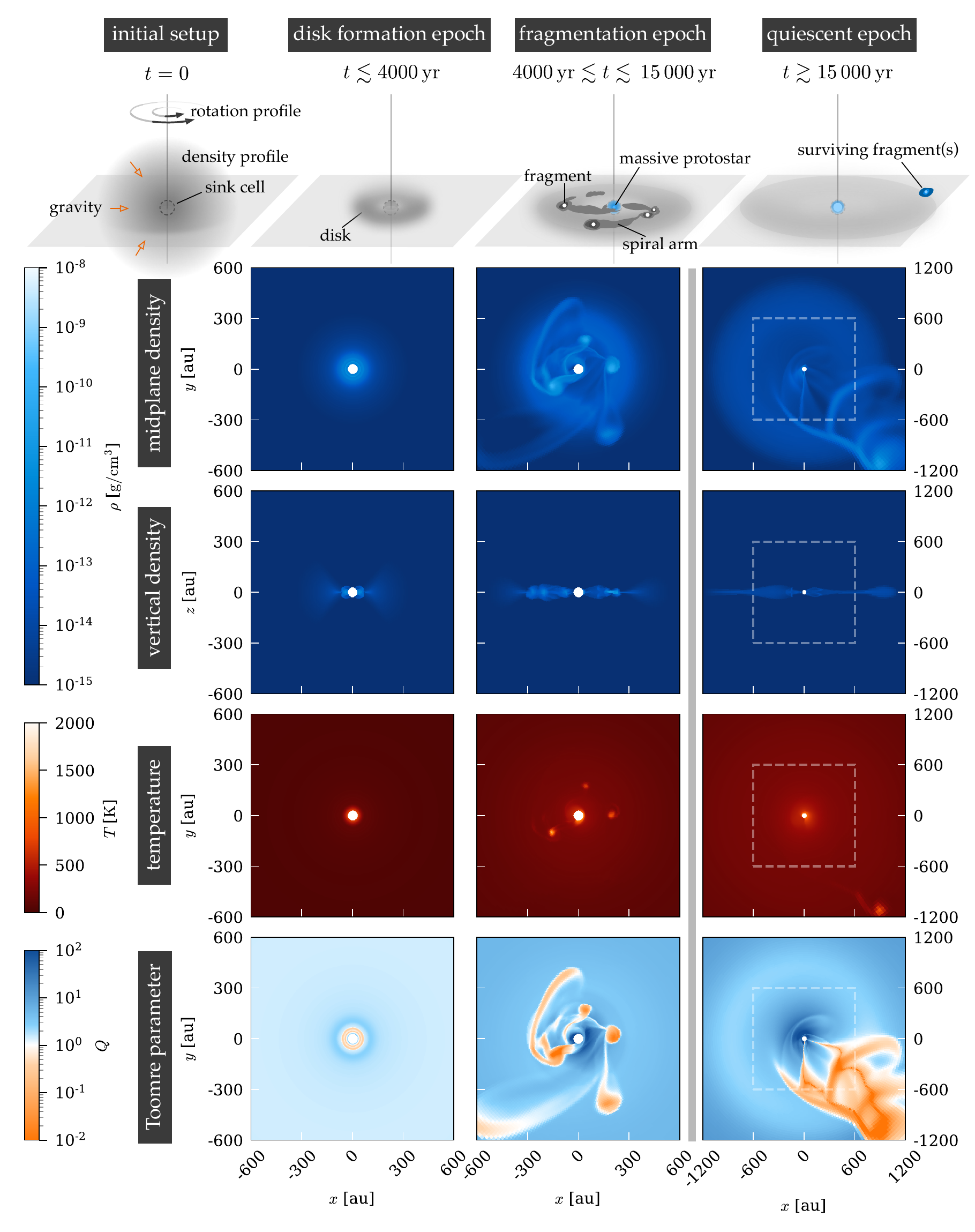}
	\caption{Time evolution of the system (run x8). In the elaboration of the plots, snapshots of the simulation were taken specifically at $3.0$, $8.0$ and $17.5\unit{kyr}$, for each respective column. The rightmost column shows double the size of the other two (the dotted square corresponds to the area shown in the snapshots located to the left). For the rightmost column, notice the existence of a surviving fragment at $\vec r \sim(900,-1200)\unit{au}$ for the midplane views.}
	\label{F: epochs}
	\end{figure*}

		Figure \ref{F: epochs} shows an overview of the time evolution of the system (for run x8), with maps of the density in the midplane ($ z = 0 $), density in the $ xz $-plane to give an idea of the vertical structure of the system, the temperature and the Toomre parameter (which will be introduced in Sect. \ref{S: method-disk-prop}).

		As soon as the simulation starts, the cloud begins to collapse. Matter free falls into the center of the cloud, where the sink cell is located, eventually forming a massive star there. Simultaneously, angular momentum conservation yields an initially axially-symmetric accretion disk that grows in size over time.

		At around $ 4 \unit{kyr} $, when the disk is about $ 200 \unit{au} $ in size, the axial symmetry is broken: spiral arms appear, and then, the disk fragments. The fragments form, interact with the disk, spiral arms and other fragments, and can get destroyed. Very roughly, fragments have densities of more than $ \sim 10^{-11} \unit{g/cm^3} $ and temperatures higher than $\sim 600 \unit{K}$; while spiral arms and other filamentary structures have densities higher than  $\sim 10^{-13}\unit{g/cm^3}$. The background accretion disk has densities higher than $10^{-15} \unit{g/cm^3}$. As shown in the density maps of Fig. \ref{F: epochs}, the whole disk (including spiral arms) is relatively thin; the pressure scale height is shown later in Fig. \ref{F: resolution-comparison}. The spiral arms and filaments are dynamic: they continuously change shape, form and merge in the course of an orbit, and they extend throughout the disk. Fragments are typically connected to each other and to the central massive protostar via spiral arms and filaments. Some fragments have the potential to form companion stars, as we describe in Sect. \ref{S: companion}. Fragments can also be surrounded by secondary disks with their own spiral arms that can lead to the formation of more fragments. These mechanisms are described in Sect. \ref{S: fragm-formation}.

	At around $ 15\unit{kyr}$, the disk stabilizes due to the increase in mass of the central massive protostar and the effects of stellar irradiation in the innermost part of the disk, stopping fragmentation of the primary disk until the end of the simulated time. We refer to this period as the quiescent epoch. Some fragments survive the fragmentation epoch, but the majority of fragments migrate toward the central massive protostar or get accreted by it. Run x8 ends at $t = 0.52 t_\text{ff}$, where $t_\text{ff} \sim \sqrt{R^3/(GM)}$ is the free fall time of the cloud; run x16 ends at $t = 0.40 t_\text{ff}$.
	
	Interestingly, the final state of run x8 is reminiscent to the young protostellar object G11.92–0.61 MM 1 reported in \cite{2018ApJ...869L..24I}. This similarity was not intended a priori in the simulation setup. We compare the outcome of run x8 and the properties of the fragment produced to this potential observational counterpart in Sect. \ref{S: comp-obs-x8}.

%\subsection{Central massive protostar}

 \begin{figure*}
 \centering
 	\setlength{\subfiglen}{0.33\textwidth}
	\subcaptionbox{}[\subfiglen]{
			\includegraphics[width=\subfiglen]{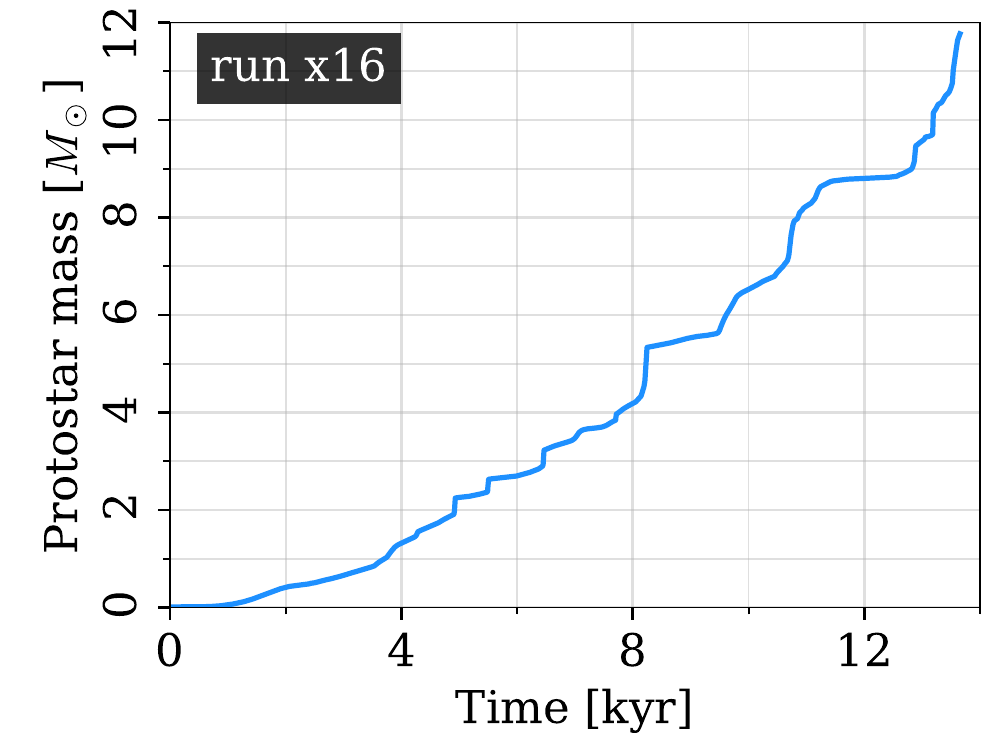}
		}
	\subcaptionbox{}[\subfiglen]{
			\includegraphics[width=\subfiglen]{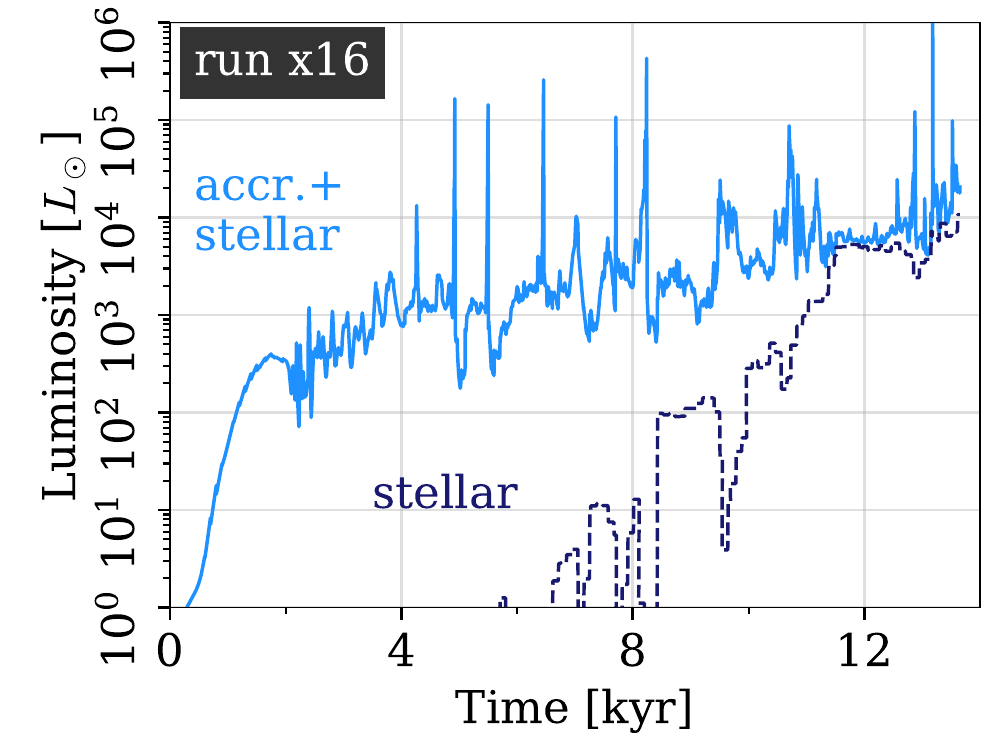}
		}
	\subcaptionbox{}[\subfiglen]{
			\includegraphics[width=\subfiglen]{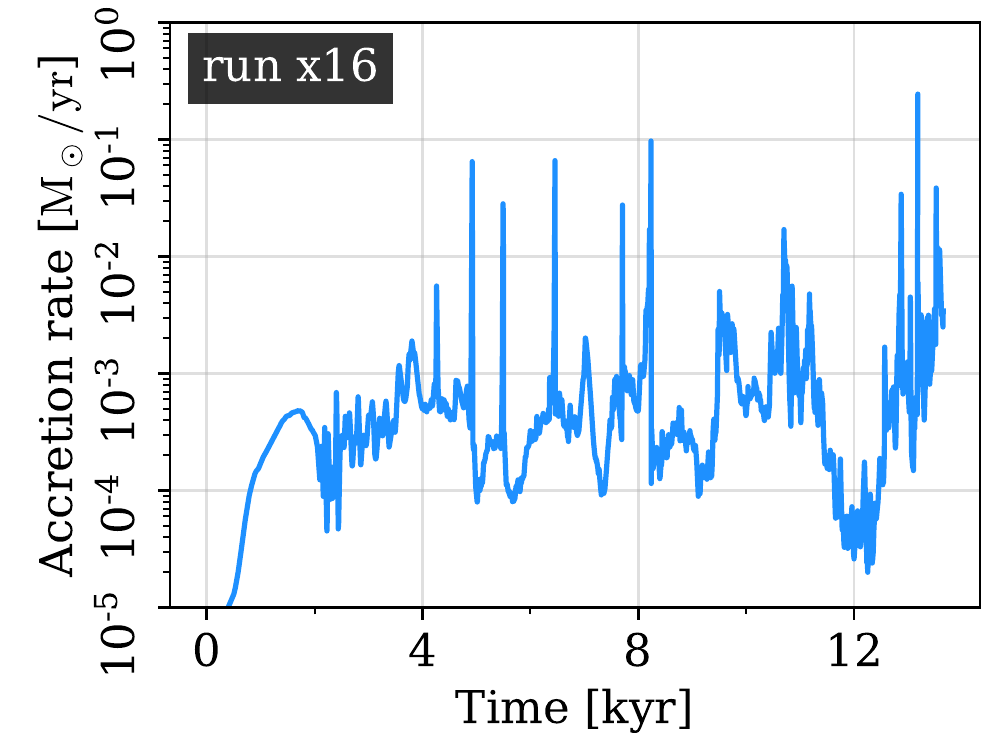}
		}
		\caption{(a) Mass, (b) luminosity and (c) accretion rate of the central massive protostar, as formed in run x16. The dotted line in panel (b) that corresponds to the stellar luminosity, is calculated using the evolutionary tracks of \cite{2009ApJ...691..823H}.}
		\label{F: starprop-x16}
\end{figure*}

	The central massive protostar formed by accretion from the disk is about $12\unit{M_\odot}$ after $\sim 14\unit{kyr}$ of evolution, as shown in Fig. \ref{F: starprop-x16}a. The total luminosity of the protostar is calculated by summing the luminosity predicted by the evolutionary tracks of \cite{2009ApJ...691..823H}, and the accretion luminosity, that is, the gravitational energy released by accretion, and computed with $L_\text{acc} = GM_\star \dot M / (2R_\star)$, where $\dot M$ is the accretion rate onto the (proto)star (Fig. \ref{F: starprop-x16}c). The total luminosity is shown in Fig. \ref{F: starprop-x16}b, together with the stellar luminosity component. At around $12 \unit{kyr}$, the massive protostar starts burning hydrogen, and the stellar luminosity becomes comparable and later dominates over the accretion luminosity. Prior to that, the luminosity of the central massive protostar is dominated by accretion, which causes accretion-driven bursts, that can be seen in Fig. \ref{F: starprop-x16}b and are discussed in Sects. \ref{S: accb} and \ref{S: impl-accb}.

% =====================================================
\section{Post-processing methods} \label{S: pp-method}

Since fragmentation was studied without the use of sink particles, sophisticated post-processing algorithms had to be developed in order to track their properties in time, which we present in this section, together with strategies for isolating the properties of the background disk.

\subsection{Properties of the disk} \label{S: method-disk-prop}
	During the fragmentation epoch, the density maps of Fig. \ref{F: epochs} show three distinct components (background disk, spiral arms and fragments), that we mentioned in the preceding section. In order to study the properties of the background disk, the spiral arms and fragments have to be filtered out from the data. We do this with a combination of the two following methods. First, in order to get the radial profile of a quantity $q$ measured for the background disk in the midplane, we take the median of all values along the azimuthal direction, i.e., for each discretized radial distance $r_i$,
		\begin{equation} q_\text{bg, midpl}(r_i) = \underset{k}{\mathrm{median}}\, q(r_i,\theta=\pi/2, \phi_k) \end{equation}
	Taking the median filters out most of the variability caused by the presence of fragments and spiral arms. For some radial profiles, an average of the radial profile in some interval of time is also used in order to eliminate strong but short-term variations. Since fragments and spiral arms are highly dynamic, a time average filters out sudden changes and yields the long-term behavior of the background disk. By using this method, one can isolate, for example, the density and temperature profiles of the primary disk, which we present and discuss in Sect. \ref{S: properties-disk}. The specific time intervals used for the average are specified in the caption of the relevant plots.

	In order to determine if the formed disk is Keplerian, the deviation from gravito-centrifugal equilibrium, or Keplerianity, of the background disk is calculated as the relative difference between the angular velocity $\Omega$ (taken for the background disk in the midplane) and its Keplerian value, $\Omega_K$:
	\begin{equation}
		\mathrm{Keplerianity} = \frac{\Omega - \Omega_K}{\Omega_K}, \text{ where }\Omega_K (r) = \sqrt{\frac{GM_\text{encl}(r)}{r^3}}
	\end{equation}
is the Keplerian angular velocity, and $M_\text{encl}(r)$ is the total mass enclosed in a sphere of radius $r$; that is, the sum of the mass of the central massive protostar and the portion of the disk enclosed. The Keplerianity was used to define the disk radius, as mentioned in Sect. \ref{S: properties-disk}.

	The accretion rate into the central massive protostar, as well as its mass, are calculated by integrating the incoming mass from all directions at the surface of the sink cell. One general limitation of the sink cell approach is the inability to distinguish between accretion of fragments onto the stellar surface and formation of close companions. This problem is discussed in Sect. \ref{S: companion}, as well as the implications for the values obtained with the method described here. 
	
	We remind the reader that no artificial viscosity is introduced in these simulations to compensate for unresolved physics; self-gravity, instead is the process that provides angular momentum transport. This motivates the idea of calculating how the mass is transported by the background primary disk, i.e., $\dot M(r)$. We take the background disk density $\rho_\text{bg}(r_i,\theta_j)$ and radial velocity $v_{r, \text{bg}}(r_i,\theta_j)$ and then we integrate in a spherical shell according to
		\[ \dot{M}(r_{i} ) = - \sum_{j} \rho_\text{bg}(r_i,\theta_j) v_{r,\text{bg}}(r_i,\theta_j) r_{i}^{2}\sin(\theta_{j}) \cdot 2\pi \cdot \Delta \theta_{j}  \]
Outside of the disk, this quantity simply expresses the mass flux transported by the infalling large-scale envelope.

In order to quantify gravitational instabilities, we use the Toomre parameter $Q$ \citep{1964ApJ...139.1217T}. This parameter measures gravitational stability against small perturbations in a rotating, self-gravitating disk, and it is defined as 
	\begin{equation}
		Q = \frac{c_s \Omega}{\pi G \Sigma}
	\end{equation}
where $\Sigma = \int \rho\, dz$ is the surface density and $c_s$ is the speed of sound; $Q<1$ indicates instability.

\subsection{Fragment detection}

	In order to study the different properties of the fragments, one must first clearly define what a fragment is. One possibility is to use the fact that fragments appear as hot points in comparison with the background disk and spiral arms (see, for example, the temperature plot during the fragmentation epoch in Fig. \ref{F: epochs}). In comparison, the density data for the same time shows spiral arms, filaments and other structures that make the definition of a fragment more difficult. In other words, spiral arms generally have  temperatures that are similar to the background disk, in sharp contrast to the fragments, allowing us to fix a criterion for fragment detection. Detecting a fragment, then, reduces to a correct detection of maxima in the temperature data for the midplane. Fragments are expected to gravitationally collapse, and therefore increase their temperature, if they have a possibility to form stars (later, we show that not all fragments can evolve further into stars).

	In order to filter out the spiral arms and the disk, a Gaussian filter is first applied to the temperature data (this blurs out the spiral arms), and then, the radial disk temperature profile is subtracted from the temperature data. After that, a temperature threshold is set so that the fragments are detected ($\sim 400 \unit{K}$). The parameters for such filters were manually tuned by selecting the values that produce fewer false positives and do not leave fragments undetected. This process produces ``fragment candidates''. It should be noted, however, that the data contains cold high-density regions that are not considered as fragments unless they overcome the temperature detection threshold. The fragment candidates are also checked by eye to filter out any remaining false positives that, when compared against the density data, turn out to be hot areas of spiral arms and not clump-like structures.

\subsection{Fragment tracking}

 	We are interested in calculating the properties of the fragments, and their evolution in time; so, we require a way to track them, using the data snapshots that are outputted every $10\unit{yr}$ of evolution, which is a much smaller timescale than the duration of one orbit ($\gtrsim 500 \unit{yr}$ at $\sim 100\unit{au}$). This is done as follows: for every fragment, the next predicted position on its orbit is estimated for a time $t_\text{current} + \Delta t$ with $\Delta \phi \sim v_\phi \Delta t$. Then, time is advanced by $\Delta t$, and maxima in the temperature are searched in a region surrounding the predicted position. When a matching fragment is found in the predicted region, a connection is made and it is registered as a fragment with a unique identification number. As a warning to the reader, even though the identification number for new fragments increases with time, it cannot be used as an indication of the total number of fragments present, since manual corrections to the output of our algorithm cause missing identification numbers.

 \subsection{Properties of the fragments}

 \begin{figure}

 \includegraphics[width=0.9\columnwidth]{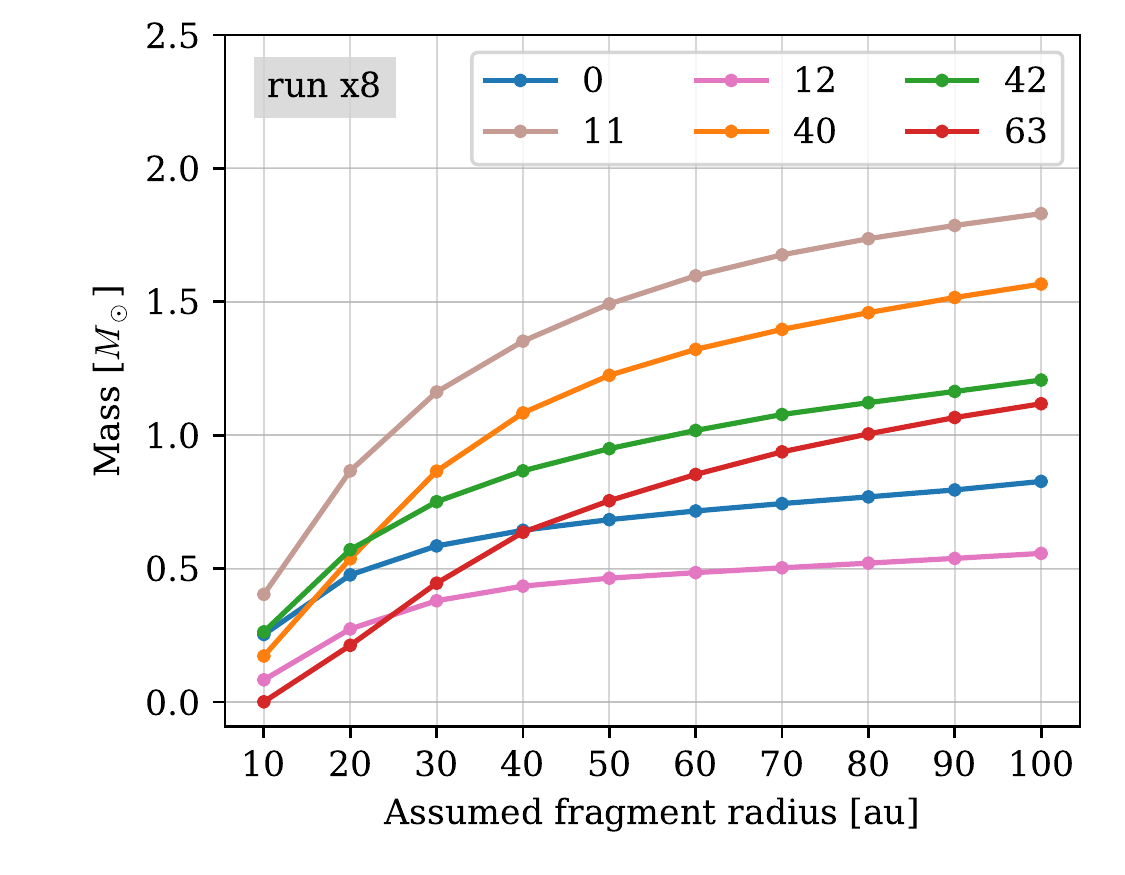}
 \caption{Variation of the time-averaged mass of a sample of fragments as a function of the assumed radius taken for integration.}
 \label{F: mass-assumed-r}

 \end{figure}

	Once the trajectory of the fragments is known, we can easily track other properties like the central temperature or the mass of the fragments.

	The central temperature is the maximum value of the temperature field of a three-dimensional region that contains the fragment. The mass of the fragment $M_\text{fragm}$ is obtained by integrating the density in a spherical region of a fixed radius (all fragments are assumed to have the same radius). A first guesstimate from the x8 density maps is $R_\text{fragm} \sim 50\unit{au}$.
	
	In order to have an idea for the upper limit of the radius of a fragment, we can estimate the order-of-magnitude of this radius assuming hydrostatic equilibrium (pressure of self-gravity $P\sim GM_\text{fragm}^2/R_\text{fragm}^4$ is equal to the pressure of the ideal gas $P \sim \langle n \rangle k_B T$), using a mean number density $\langle n\rangle \sim 1.5 \cdot 10^{17} \unit{cm^{-1}}$, extracted from the density data, and a temperature of $2000\unit{K}$. This yields $R_\text{fragm} \sim 30 \unit{au}$ for a fragment of $1\unit{M_\odot}$, and $R \sim 50\unit{au}$ for a fragment of $2\unit{M_\odot}$.

	 To justify our choice, we performed mass calculations with several assumed radii (the results are shown in Fig. \ref{F: mass-assumed-r}). Small assumed radii give a steep gradient in the mass value, suggesting that most of the mass of the fragment has not yet been enclosed in the region of integration. At around $50\unit{au}$ for run x8, the curve shows a more stable value. Bigger radii would not only enclose the fragment, but also parts of the disk and surrounding spiral arms. A similar analysis done for run x16 yields a radius of $40\unit{au}$. These considerations mean that the mass of the fragments reported here have an uncertainty of around 20\%.

% =========================================
\section{The fragmenting accretion disk} \label{S: accretion disk}

	\begin{figure*}
		\subcaptionbox{Keplerianity (midplane)}[0.33\textwidth]{
			\includegraphics[width=0.33\textwidth]{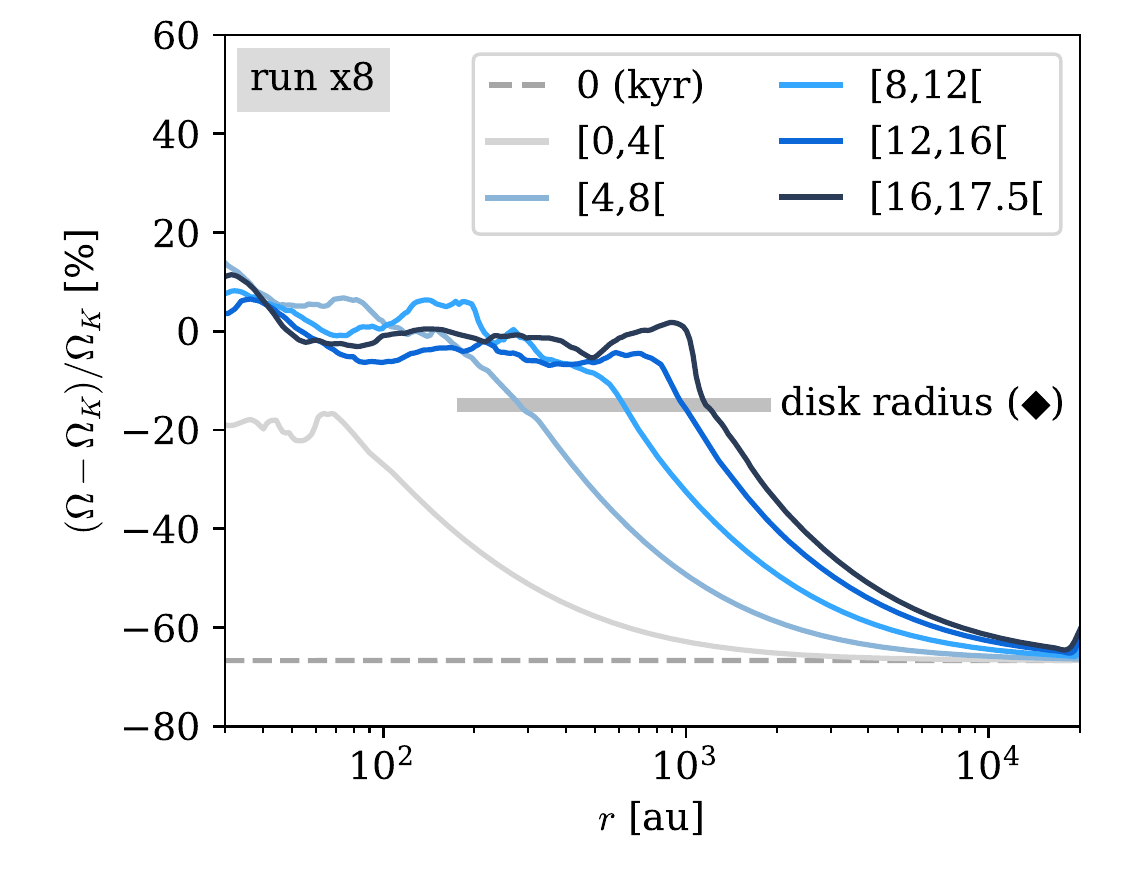}
		}
		\subcaptionbox{Density (midplane)}[0.33\textwidth]{
			\includegraphics[width=0.33\textwidth]{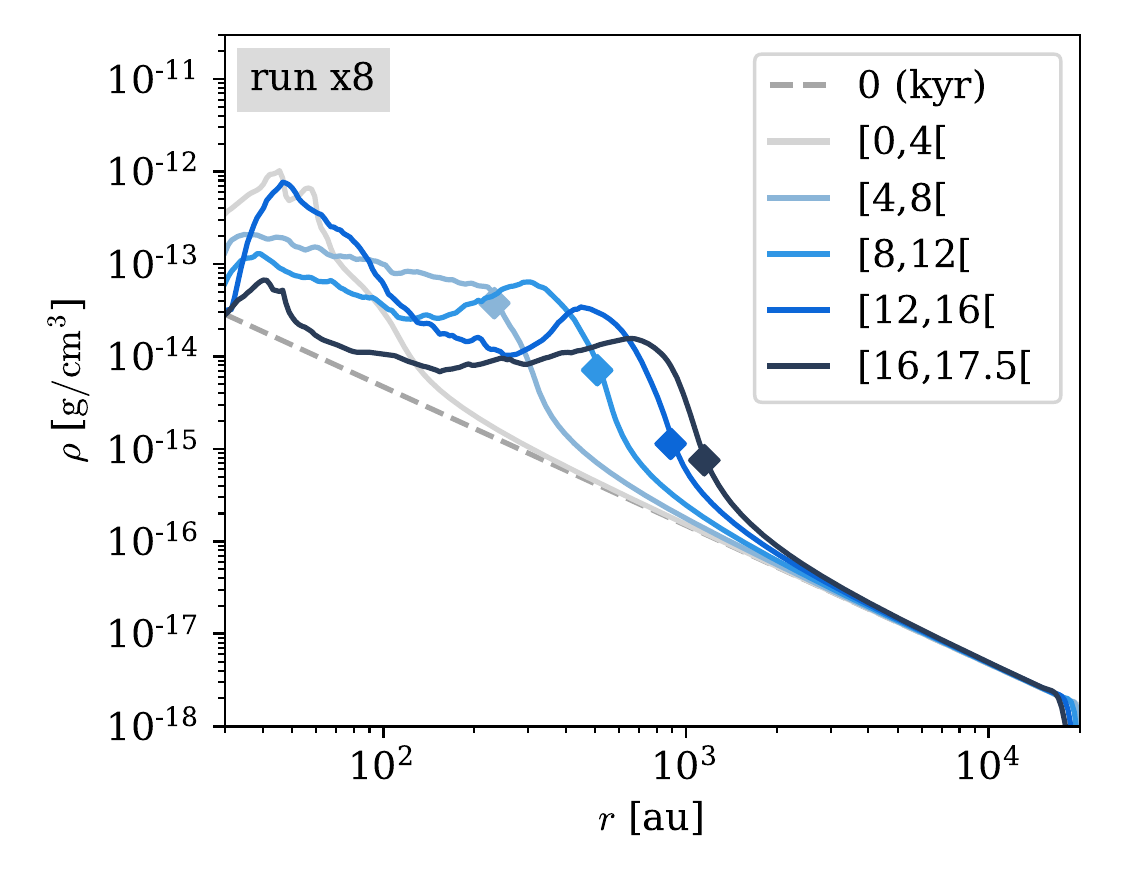}
		}
		\subcaptionbox{Surface density}[0.33\textwidth]{
			\includegraphics[width=0.33\textwidth]{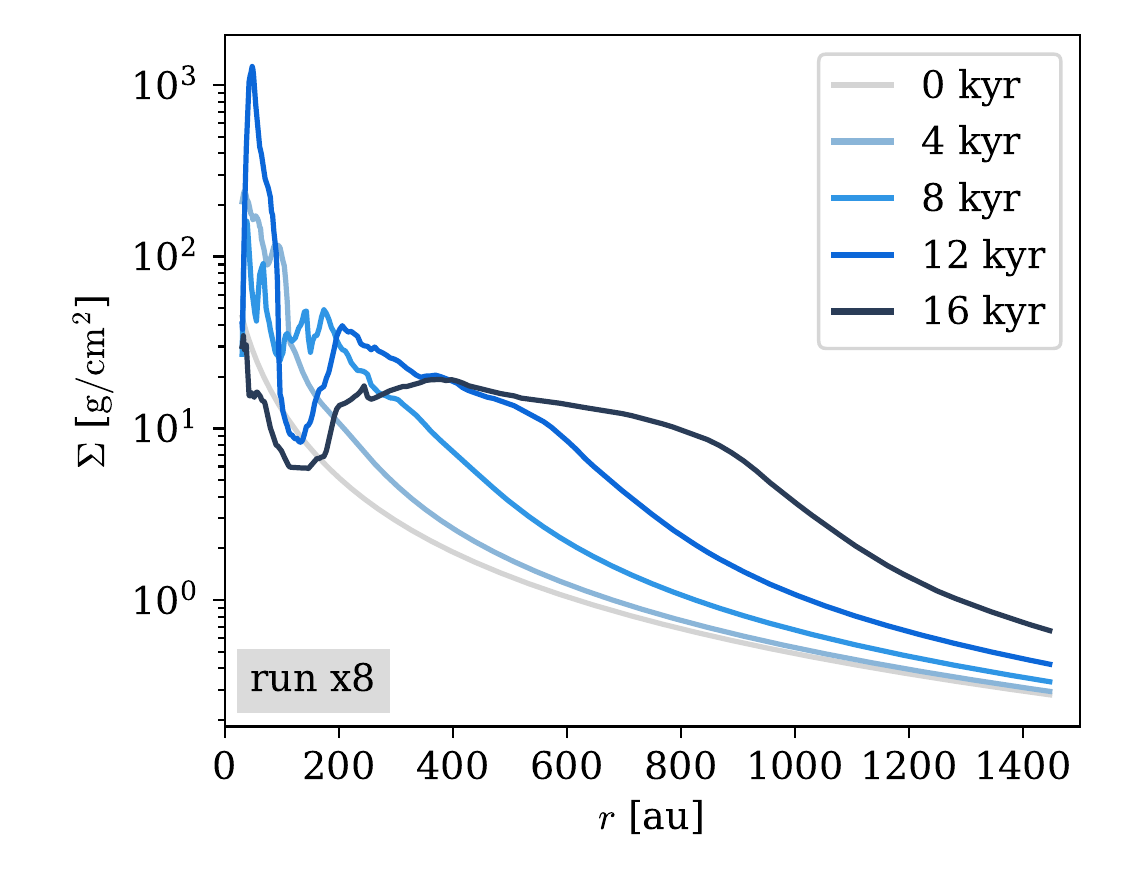}
		}
		\subcaptionbox{Temperature (midplane)}[0.33\textwidth]{
			\includegraphics[width=0.33\textwidth]{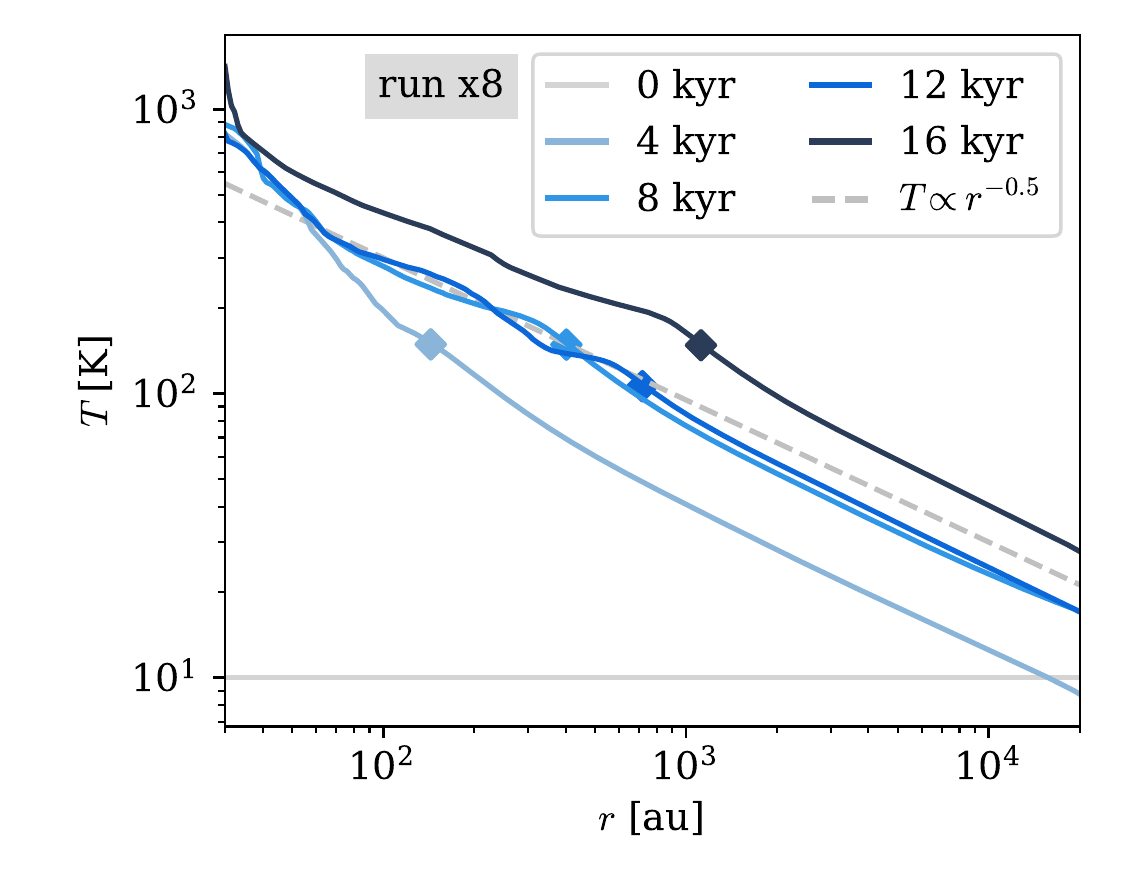}
		}
		\subcaptionbox{Radial mass flow (sph. shell)}[0.33\textwidth]{
			\includegraphics[width=0.33\textwidth]{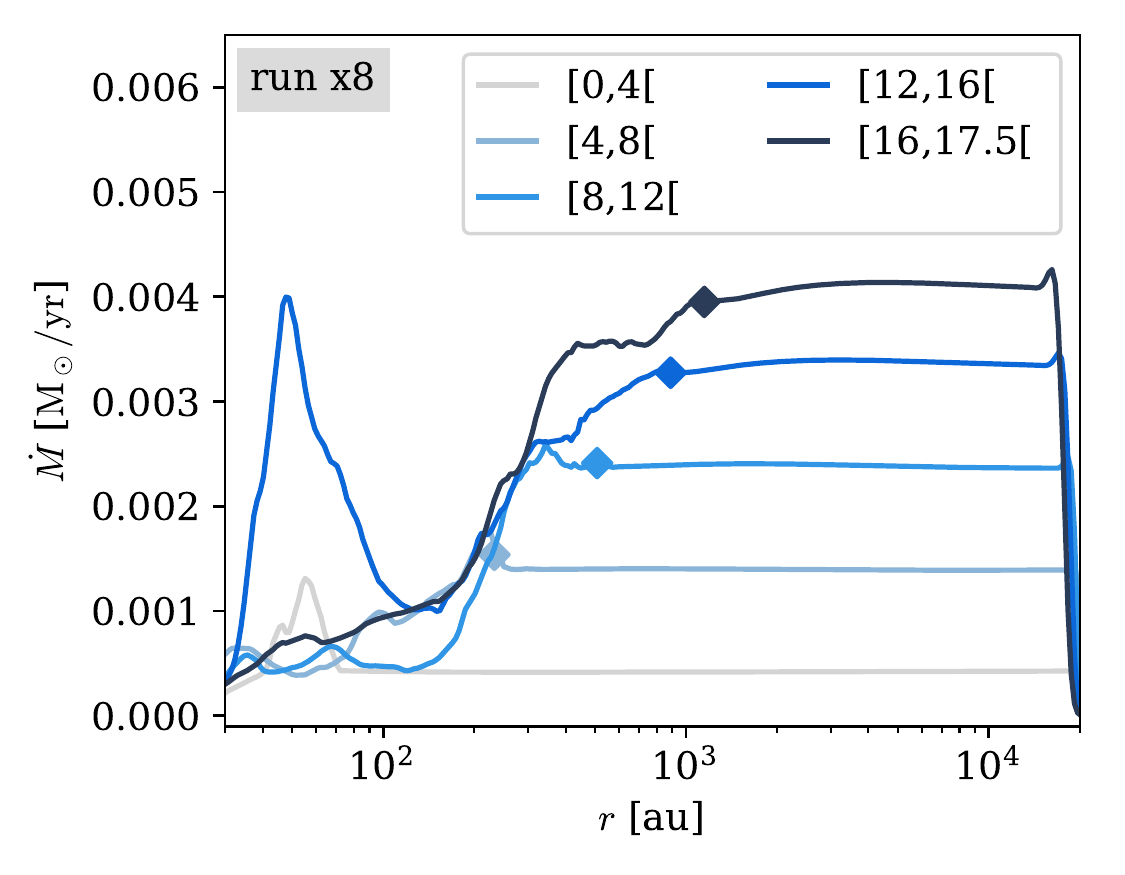}
		}
		\subcaptionbox{Accretion rate}[0.33\textwidth]{
			\includegraphics[width=0.33\textwidth]{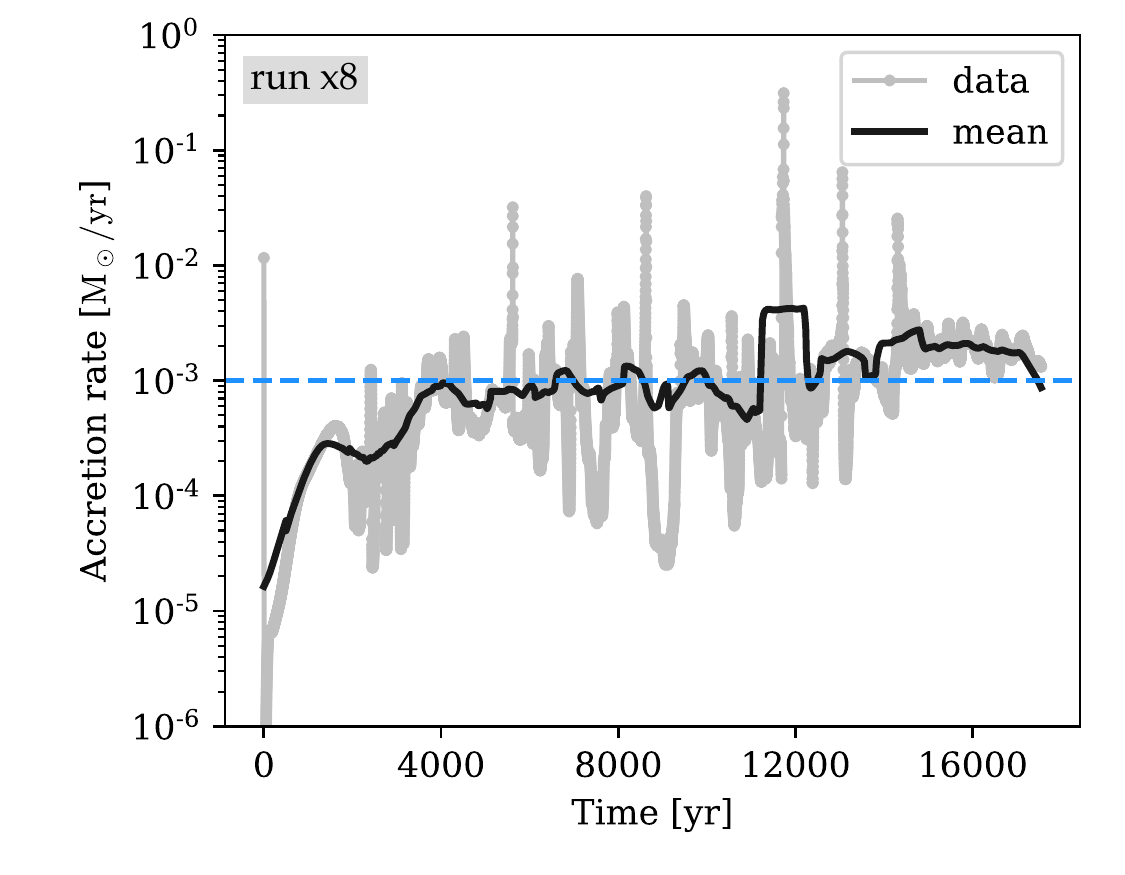}
		}

		\caption{(a-e) Radial profiles of several quantities for the background disk. The intervals in profiles (a), (d) and (e) mean that the quantity has been time-averaged in that interval, excluding the endpoint. The filled diamond indicates the value at the disk radius. (f) Accretion rate into the central massive protostar as a function of time.}
		\label{F: profiles}
	\end{figure*}
	
The next sections present the results of the highest resolution simulations, and an in-depth analysis of the internal processes that govern fragmentation in the disk.

\subsection{Properties of the disk} \label{S: properties-disk}
	Radial profiles of several quantities related to the background disk are presented in Fig. \ref{F: profiles}, using run x8. The density, Keplerianity and radial mass flow are averaged in time, as discussed in Sect. \ref{S: pp-method}. 
	
	After the disk formation epoch (from 4\,kyr until the end of the simulation), the disk is approximately Keplerian, as shown by the Keplerianity profile, with a variability of $\sim 10$\%. We used this fact to define the disk radius as the point in which the Keplerianity of the background disk drops below $-15$\%. For larger scales, the low values of Keplerianity indicate infall from the large-scale envelope, which replenishes the disk. The initial rotation profile chosen, with $\beta_\Omega=-3/4$, means that the gas is uniformly non-Keplerian, and during the disk formation epoch (0 to 4\,kyr), the disk builds up until it reaches gravito-centrifugal equilibrium. In some of the panels, the disk radius is indicated as a reference with filled diamonds.

	The density profiles show that the background disk stays mainly between $10^{-15}$ and $10^{-13} \unit{g/cm^3}$. A higher density region is observed near the edge of the disk, and corresponds to a centrifugal barrier.
	
	As expected, the temperature profile of the collapsing cloud, including the accretion disk, increases over time, although we observe that the profile increases more slowly during the fragmentation epoch. The temperature profile is approximately proportional to $r^{-0.5}$. At the late stages of the simulation (cf. the line for 16 kyr in Fig. \ref{F: profiles}c), there is an increase in temperature in the inner parts of the disk ($\lesssim 40\unit{au}$), probably due to stellar irradiation, since the central massive protostar has just begun burning hydrogen at that time.
	
	 The radial mass flow transported by the background disk is different from the radial flow in the infalling large-scale envelope. This shows that mass from the infalling envelope is not only being transported onto the central massive protostar, but it is also deposited into the fragments and spiral arms, and that most of the mass is delivered to the massive protostar by means of the spiral arms and fragment accretion, and not through the background disk, during the fragmentation epoch.

	Data from run x8 presents a region with a seemingly unusual overdensity (more associated with the densities of spiral arms) at about 12 kyr in the region $r \lesssim 100\unit{au}$, which is particularly evident on the surface density and radial mass flow profiles. As a result of an accretion event in run x8, the inner disk gets temporarily a higher density ``ring like'' accretion structure. As a consequence, computing the background disk density as the median value in an annulus yields a higher density. Such accretion structures are also ocasionally observed in the other runs, although with shorter durations. Disregarding this short-term feature, the general trends shown in Fig. \ref{F: profiles} are also observed in the other runs.

	The accretion rate onto the central massive protostar as a function of time shows great variability, since fragmentation creates small accretion events on top of of having a very smooth, constant process, as described in Sects. \ref{S: accb} and \ref{S: impl-accb}. Despite that, the mean accretion rate in the fragmentation epoch is fairly constant, at about $ 10^{-3}\unit{M_\odot/yr}$, although a slight increment with time in the accretion rate is observed in both panels (e) and (f), due to simple acceleration of the gas in free fall.

\subsection{Local vs global Toomre parameter}
	\begin{figure}
		\includegraphics[width=\columnwidth]{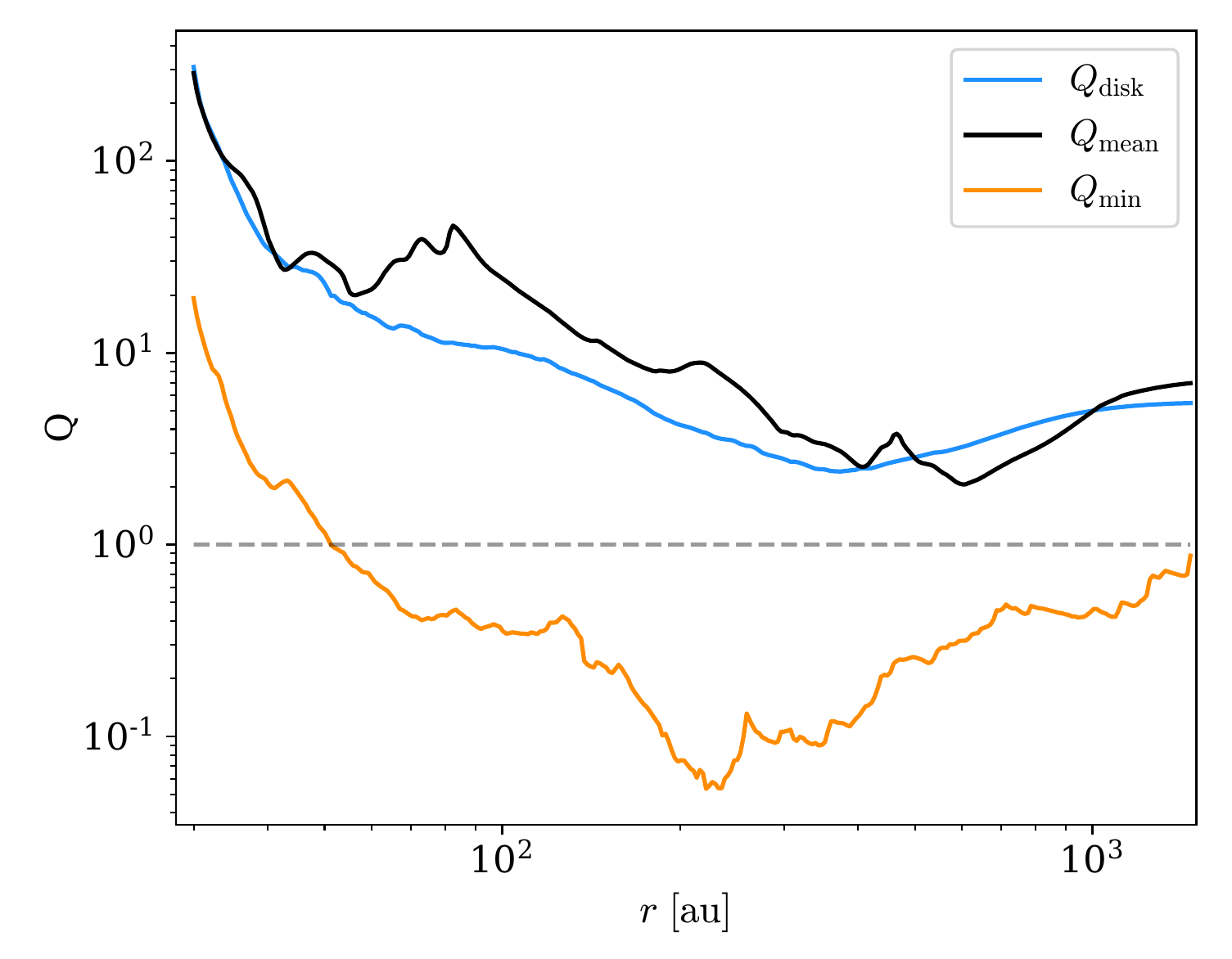}
		\caption{Comparison of the Toomre parameter values computed for the background disk ($Q_\text{disk}$), the mean value ($Q_\text{mean}$) and the minimum values corresponding to the effects of the fragments and spiral arms (corresponding to $Q_\text{min}$). These curves were calculated with run x16, and they are time-averaged for the fragmentation epoch.}
		\label{F: Q_local_global}
	\end{figure}

	The Toomre maps in Fig. \ref{F: epochs} show that the Toomre parameter is highly dependent on position and time. In Fig. \ref{F: Q_local_global}, we present the Toomre parameter as a function of radius, averaged in time for the fragmentation epoch. The curve for $Q_\text{disk}$ is calculated by taking the azimuthal median of all the variables and then computing $Q$, and it corresponds to the values for the background disk. On the contrary, the curves for $Q_\text{mean}$ and $Q_\text{min}$ are obtained by first calculating $Q(r,\phi)$ and then taking the mean and the minimum, respectively. $Q_\text{min}$ represents the regions of the disk that undergo most fragmentation, i.e., spiral arms and fragments, while $Q_\text{mean}$ may be interpreted as a value of $Q(r)$ obtained observationally for spatially unresolved sources (i.e., substructures such as spiral arms and fragments are not detected).
	
	The background disk is Toomre-stable, and no fragmentation is expected to occur in these regions; however, spiral arms and fragments are subject to fragmentation. The mean value, on the other hand, shows stability. The whole disk is, therefore, globally stable while being locally unstable. This means that the substructures of the disk have to be resolved accurately in order to capture fragmentation; an insufficiently resolved disk may appear to be Toomre-stable while undergoing fragmentation at unresolved scales.

\subsection{Spiral arm formation}
	In all runs, at the end of the disk formation epoch ($t \sim 400 \unit{yr} $, see Fig. \ref{F: epochs}), a ring-shaped region in the disk becomes Toomre-unstable and develops small inhomogeneities that become two primordial spiral arms at opposite sides of the disk, that is, they arise from the $m=2$ mode described in \cite{1996ApJ...456..279L} \citep[see also, e.g.,][]{2008ApJ...681..375K, 2011ApJ...732...20K}. The first fragments form at the outermost parts of the spiral arms.

	New spiral arms are created by convergent flows, as shown in Fig. \ref{F: subdisks}a; the flows left and below the fragment are convergent and mass accumulates as spiral arms.
These type of flows are frequently created by turbulent motions arising after fragment interactions.

% =========================================================
\section{Fragments}  \label{S: fragments}

\begin{figure*}
	\centering
	\includegraphics[width=\textwidth]{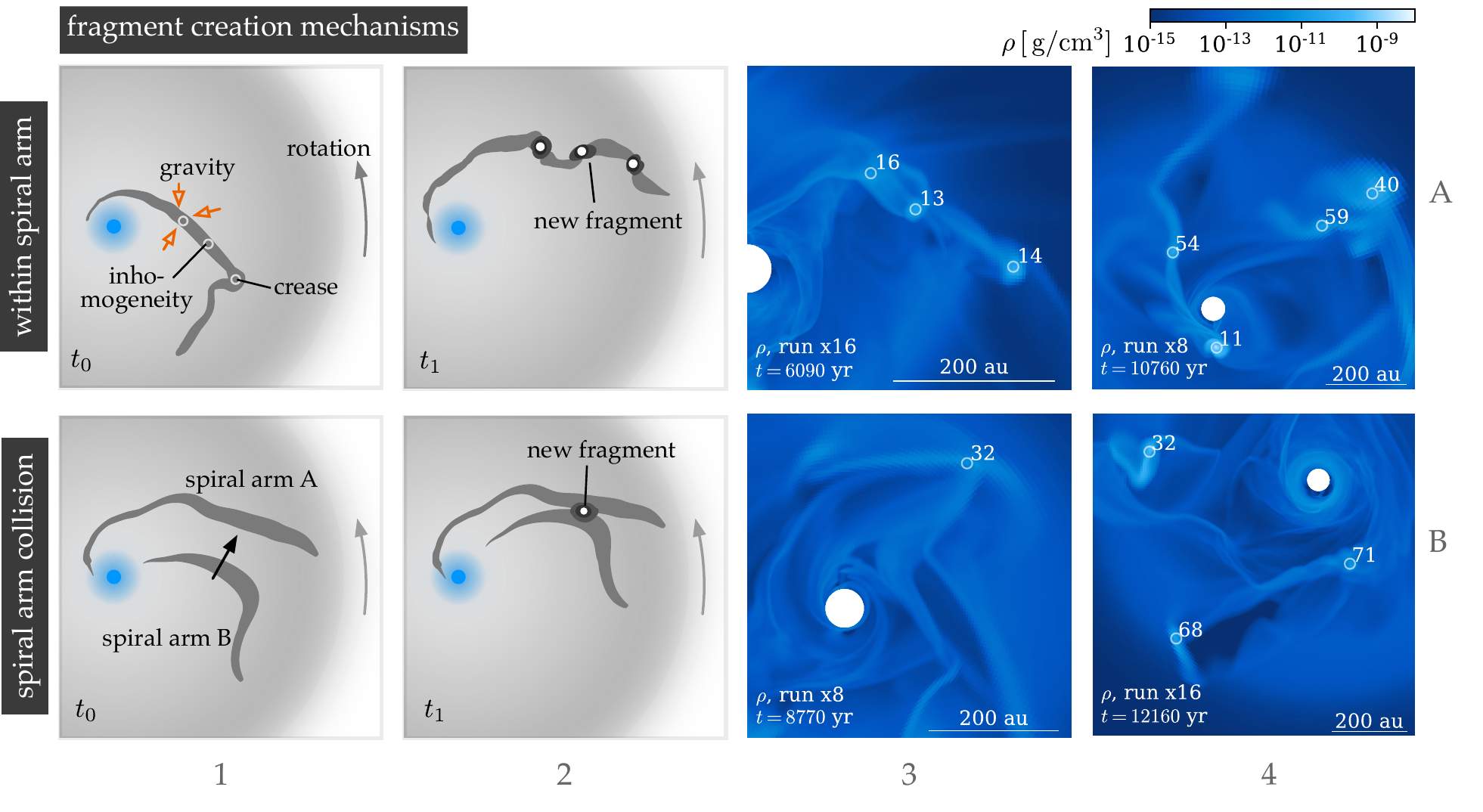}
	\caption{Fragment creation mechanisms. The color scale for the midplane density maps is the same as the one shown in Fig. \ref{F: epochs}.}
	\label{F: fragm-creat}
\end{figure*}

\subsection{Fragment formation} \label{S: fragm-formation}

 Fragments form out of one or more spiral arms. This two-step fragmentation process (first the disk forms spiral arms, and then they fragment), is in agreement to what \cite{2016MNRAS.458.3597T} reported in the context of simulations of self-gravitating protoplanetary disks. We have identified two distinct mechanisms of fragment formation in our simulations: a local collapse within a spiral arm and a formation triggered by a collision of two spiral arms. These mechanisms are present at two different scales: the primary disk, and the secondary disks that form around the fragments (we discuss the latter in more detail in Sect. \ref{S: subdisks}).

  The first process, local collapse within spiral arms, is illustrated in panels A1 and A2 of Fig. \ref{F: fragm-creat}. When small inhomogeneities are developed within a spiral arm, or when a spiral arm develops a crease, it becomes Toomre-unstable and a collapse starts, forming small fragments that deform the spiral arm. This process is exemplified by the creation of fragments 16, 13 and 14 of panel A3 (run x16), where the spiral arm develops inhomogeneities that grow into fragments. The other example is given in panel A4, where fragments 54 and 59 are formed by a previous crease; in the case of fragment 59, the crease was formed in a spiral arm of the secondary disk around  fragment 40.

  The second process, spiral arm collision, is illustrated in panels B1 and B2. When two spiral arms or regions of high density collide, they create perturbations that make the region Toomre-unstable and therefore, a collapse is triggered. The example in panel B3 is from run x8, and shows the formation of fragment 32, which is generated by this mechanism. Panel B4 provides an additional example from run x16, where two spiral arms, connecting fragments 32 and 68 to the inner disk, have collided and gave rise to fragment 71.

  Early during the fragmentation epoch, both processes occur predominantly in the primary disk. However, towards the end of the fragmentation epoch, and as fragments gain more mass, the fragmentation inside substructures becomes dominant: spiral arm breakdown and spiral arm collisions become more frequent in the secondary disks and yield fragmentation. This behavior was observed especially in runs x4 and x16. This hierarchical fragmentation is in principle analogous with the observations and conclusions offered in \cite{2015A&A...581A.119B} and \cite{2019A&A...621A.122B}.

\subsection{Number of fragments}

\begin{figure}
	\centering
	\includegraphics[width=0.95\columnwidth]{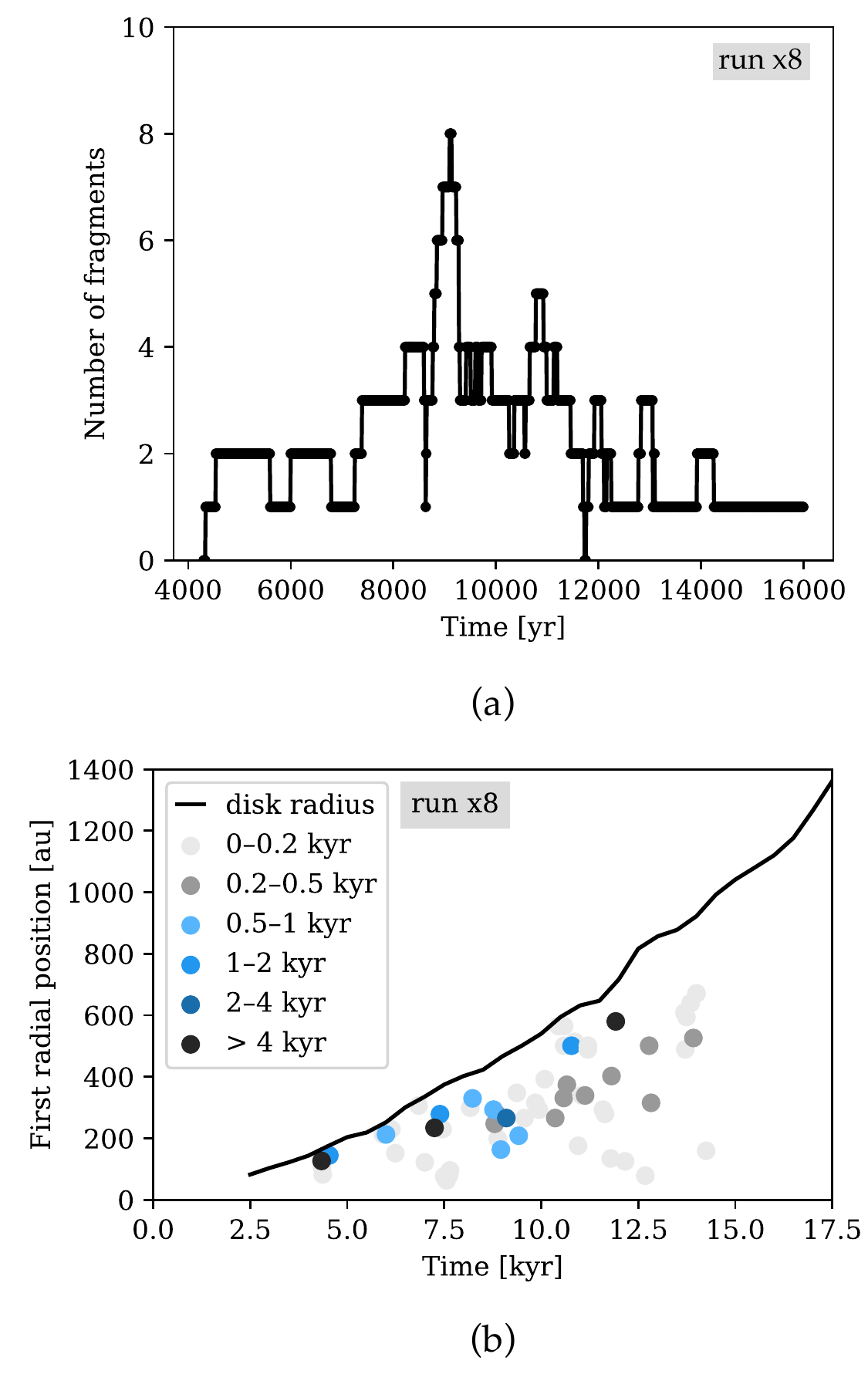}
	\caption{(a) Number of fragments with a minimum lifetime of 200 yr present in the simulation as a function of time. (b) Radial position at formation time for each fragment, color-coded with the fragment lifetime.}
	\label{F: num-fragm}
\end{figure}

	The number of fragments present in run x8 as a function of time is plotted in Fig. \ref{F: num-fragm}. We remind the reader that the numbers next to the fragments in the plots throughout this paper are mere identification numbers. In total, 60 fragments are detected during the fragmentation epoch (4--15$\unit{kyr}$). From these, 22 live longer than 200 yr; these fragments will be the focus of the analysis that follows. At a given time, there are fewer than 9 fragments present in the disk. A discussion about the number of surviving fragments is offered in Sect. \ref{S: fate}. A peak in fragmentation is seen at around 9 kyr, although we do not consider it as of great interest, since it is not observed in the other runs (see Fig. \ref{F: fragm-convergence-1}).

	Lifetimes of the fragments (color code) and the radial location when they are formed (vertical axis) are shown in Fig. \ref{F: num-fragm}b, as a function of time (horizontal axis). We see that the longest lived fragments are formed in the outer disk, and that the inner disk forms fewer fragments over time. This trend is tied to the fact that fragmentation is occurring in the spiral arms of the primary disk. The figure was generated with the data from run x8. In run x16, as the spiral arms of the secondary disks fragment, new long-lived fragments start to form again in the inner disk.

\subsection{Mass} \label{S: fragm-mass}

\begin{figure}
	\centering
	\includegraphics[width=\columnwidth]{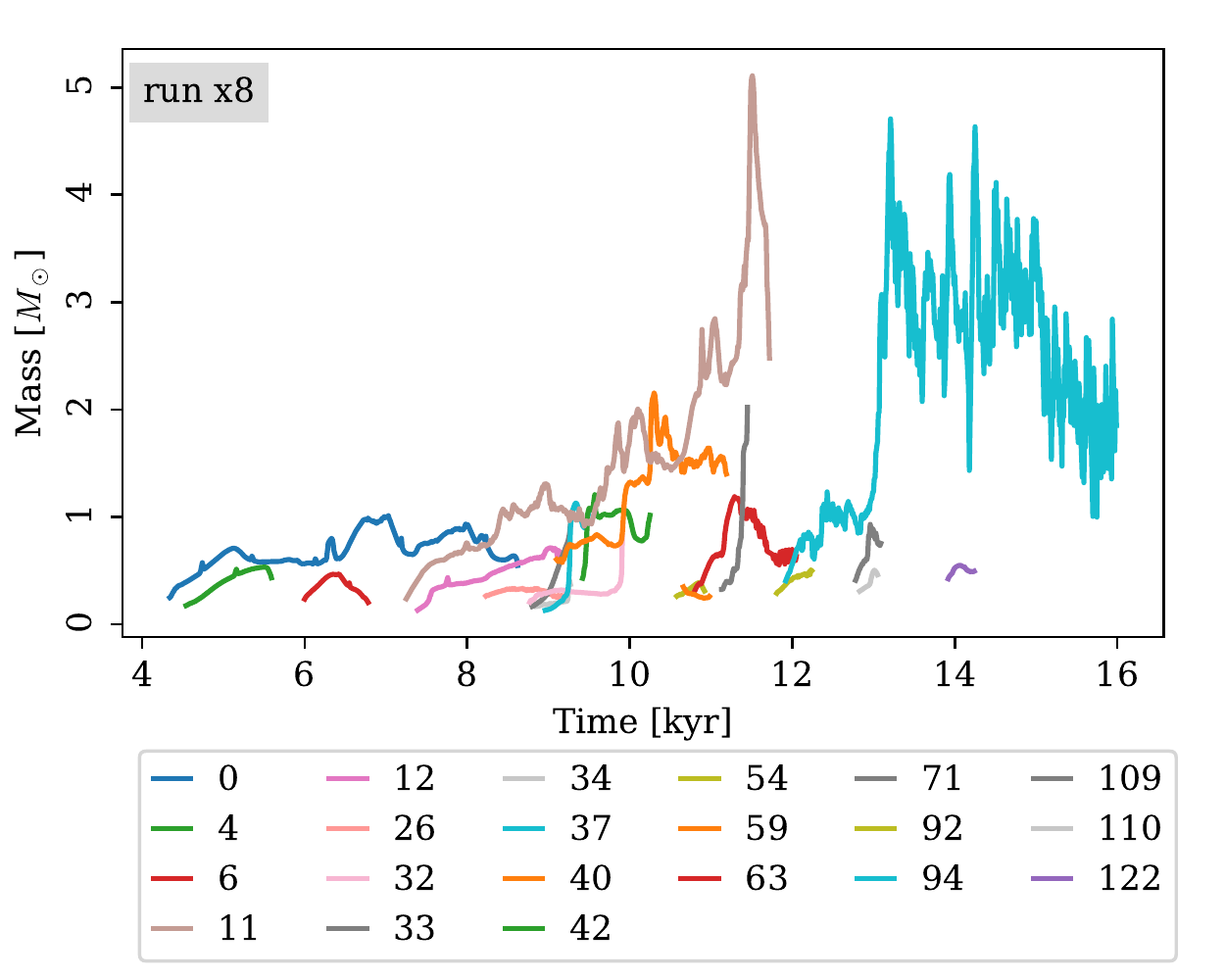}
	\caption{Mass of the fragments with life span longer than 200 yr. A fixed radius of 50 au is assumed for the calculations of this plot.}
	\label{F: fragm-mass-x8}
\end{figure}

\begin{figure*}
	\centering
	\includegraphics[width=0.9\textwidth]{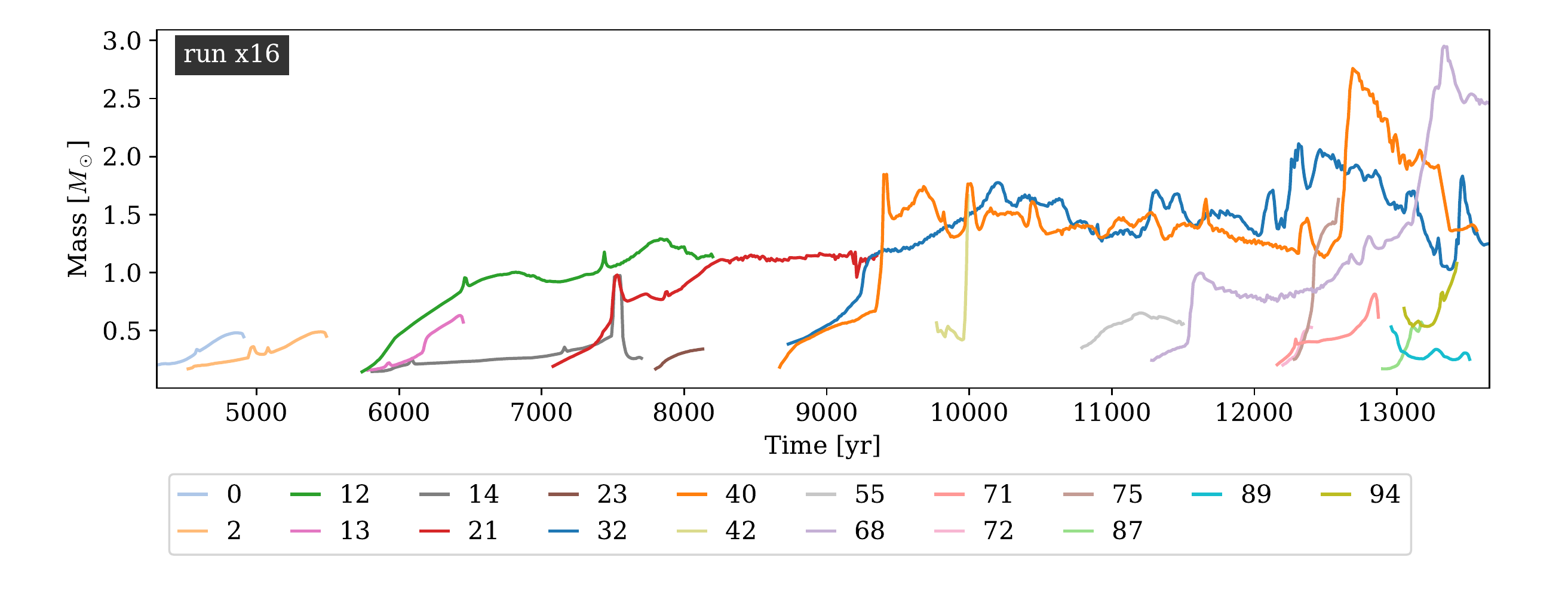}
	\caption{Masses of the fragments with life span longer than 200 yr, for run x16. A fixed radius of 40 au is assumed for the calculations of this plot.}
	\label{F: fragm-masses-x16}
\end{figure*}

	The masses of the fragments are presented in Figs. \ref{F: fragm-mass-x8} and \ref{F: fragm-masses-x16} (corresponding to runs x8 and x16, respectively), obtained using the method described in Sect. \ref{S: pp-method}. The masses of the fragments are of a few solar masses, and increase over time. Peaks and sudden increases are observed, and correspond to fragment mergers, that are described in Sect. \ref{S: interactions}. Some decreases and oscillatory behavior are caused by the action of spiral arms associated to the secondary disk, as described in Sect. \ref{S: subdisks}. The first fragments formed tend to have lower masses, compared to fragments formed later in time. This is not surprising, since the fragments form at larger radii, where larger portions of the disk become unstable, but also since the radial mass flow of the background disk increases both with the radial position as well as with time. More massive fragments tend to have longer lifetimes, as well, since they are better gravitationally bound and more immune to the fragment destruction mechanisms described in the next sections.

	Two special cases occurring in Fig. \ref{F: fragm-mass-x8} should be discussed in more detail. Fragments 11 and 94 show unusually high masses, and, in the case of fragment 94, high variability. Fragment 11 undergoes several mergers and gains mass just before it goes into the sink cell. The erratic behavior of fragment 94, in the other hand, is mostly numeric: it moves outwards in the simulation domain, where there is less resolution; and by effect of numerical diffusion, increases in radius, making our assumed radius of 50 au insufficient in the calculation. The mass of fragment 94, assuming a bigger radius of $\sim 100 \unit{au}$, is $\sim 6.5\unit{M_\odot}$ at $t = 13\unit{kyr}$.

\subsection{Hydrostatic cores}
As we discuss in Sect. \ref{S: companion}, fragments collapse and form hydrostatic cores. We remind the reader that the masses presented in this section are calculated by assuming a fixed fragment radius of $50\unit{au}$ for run x8, and $40\unit{au}$ for run x16, which means that the enclosed region includes the hydrostatic core as well as the secondary disk. This is important when comparing to models of core collapse, such as the ones described in \cite{2018A&A...618A..95B}: for an initial core mass of a few solar masses, they find hydrostatic first Larson cores of radii of $\sim 3\unit{au}$, and masses of the order of a few $10^{-2} \unit{M_\odot}$.

In run x16, cell sizes of $\lesssim 1\unit{au}$ are reached at radial positions of $r\lesssim 100\unit{au}$, allowing us to barely resolve the hydrostatic core region of fragments located there with a few grid cells, enough for an order of magnitude check with a core collapse model. We calculated the enclosed mass in the inner $\approx 3\unit{au}$ of fragment 12 (run x16) over its evolution. During the gravitational collapse of the fragment (Fig. \ref{F: vert-struct}a), the mass of the inner $\approx 3\unit{au}$ increased, until a value of a few $10^{-2} \unit{M_\odot}$ was reached, consistent with the results of \cite{2018A&A...618A..95B}. This high-density inner region is shown also in Figs. \ref{F: subdisks}b and \ref{F: vert-struct}b, surrounded by the secondary disk (the figures correspond to the midplane and vertical cuts in the density field, respectively). As a reference for the reader, the yellow circle in Fig. \ref{F: subdisks}b indicates the inner $3\unit{au}$ (roughly the size of the first core), and the yellow circle in both panels of \ref{F: vert-struct} indicates the assumed size of a fragment, used for the calculation of its (total) mass.
	
\subsection{Fragment dynamics: orbits} \label{S: fragm-orbits}

\begin{figure}
	\centering
	\subcaptionbox{}[0.48\columnwidth]{
		\includegraphics[width=0.48\columnwidth]{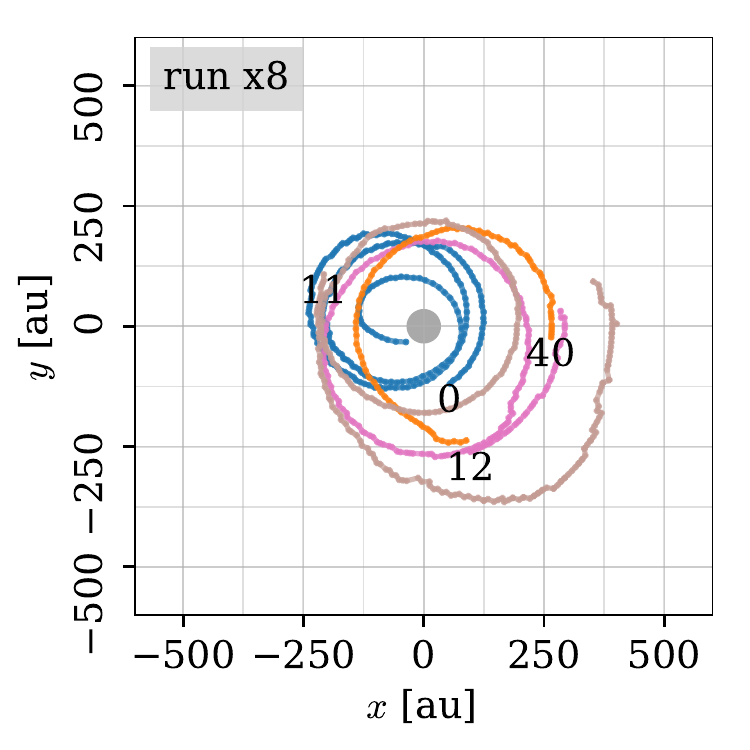}
	}
	\subcaptionbox{}[0.48\columnwidth]{
	\includegraphics[width=0.48\columnwidth]{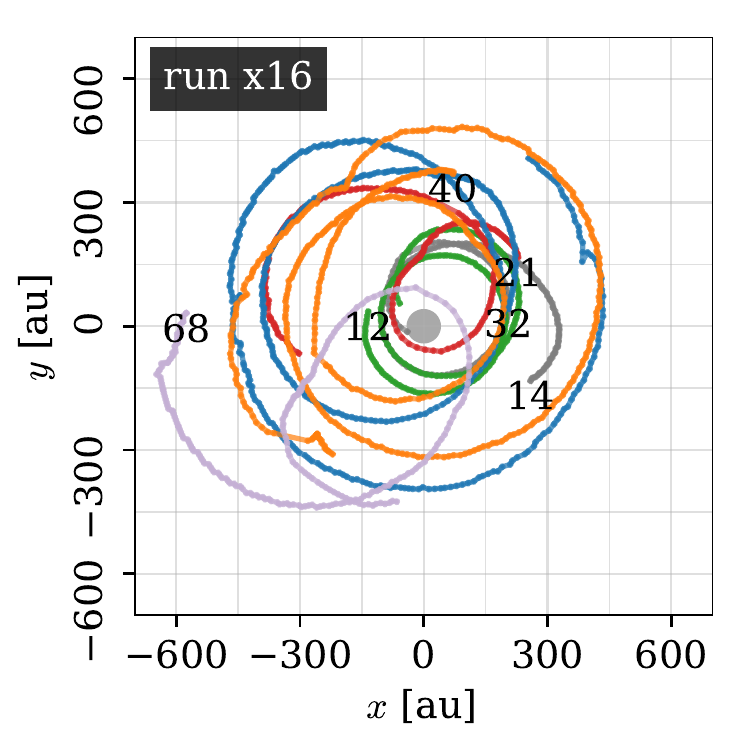}
	}
	\caption{(a) Orbits of fragments with a life longer than 1 kyr, during the time interval $[6,10] \unit{kyr}$, for run x8. (b) Orbits of all the fragments with a life longer than 1 kyr, during all the simulated time, for run x16. The fragment identification number is located at the starting point of the orbit, and the color code is the same as the one used in Figs. \ref{F: fragm-mass-x8} and \ref{F: fragm-masses-x16}.}
	\label{F: fragm-orbits}
\end{figure}

The rest of Sect. \ref{S: fragments}  will be devoted to the study of how fragments move in space, their interactions with their environment, and their substructures. First, an overview of the orbits is presented. In Fig. \ref{F: fragm-orbits}, the orbits of long lived fragments ($>1 \unit{kyr}$) have been plotted. During the simulated time period, only about 8 fragments complete more than one orbit (both runs). The orbits of the fragments are highly influenced by their interactions with spiral arms and other fragments (fast migration, mergers, etc.; more details below). They also stay in the midplane, except some orbits of run x16, that show a small inclination ($\lesssim 2^\circ$) with respect to the midplane during certain times. The average period is $\sim 1\unit{kyr}$. The orbits of the fragments are highly eccentric, with an average eccentricity of $\sim 0.5$, calculated by taking the minimum and maximum radial positions of the fragments. Fragments undergo different types of interactions that cause changes in the eccentricity of their orbits, including fast inward migration due to spiral arm action and gravitational interaction with other fragments.

The orbits of long-lived fragments 40 and 32 of run x16 (Fig. \ref{F: fragm-orbits}b), develop in the middle disk, and correspond to fragments that survive at the end of the simulated time. These fragments are also more massive than short-lived ones (according to Fig. \ref{F: fragm-masses-x16}, their masses are $\sim 1.5 \unit{M_\odot}$), and therefore their self-gravity provides more stability against interactions with the environment.

The orbits presented here, however, do not take into account the effects of the formation of second Larson cores and the fate of fragments that enter the numerical sink cell. A more detailed discussion is presented in Sect. \ref{S: fate}.

\subsection{Interactions} \label{S: interactions}

\begin{figure*}
	\centering
	\includegraphics[width=\textwidth]{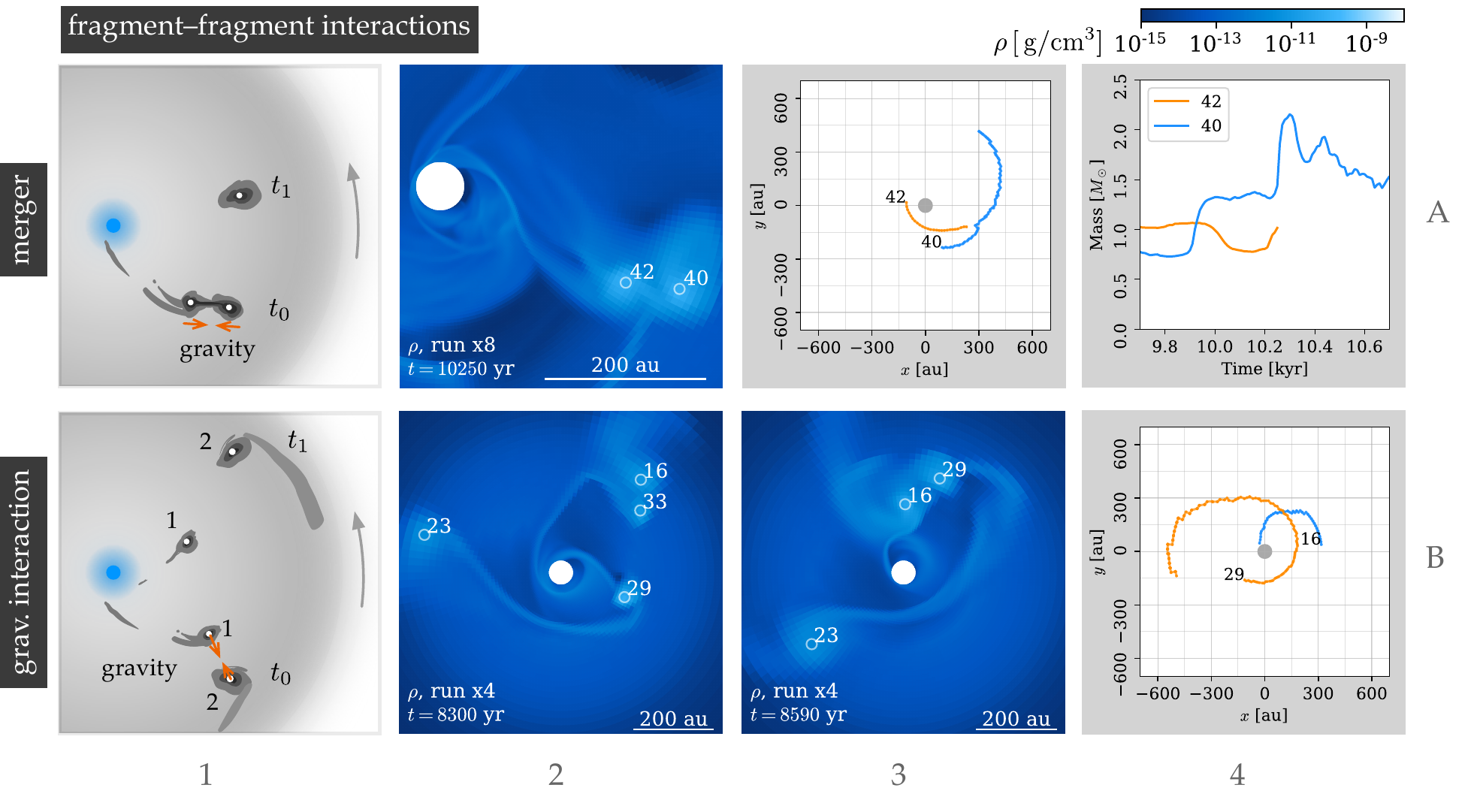}
	\caption{Fragment-fragment interactions. In panels A3 and B4, the fragment ID is shown at the starting point of the orbit. In the case of panel A3, the time window shown in the orbits is 1.3 kyr; and in the case of panel B4, 1.2 kyr. }
		\label{F: fragm-events}
\end{figure*}

\subsubsection{Fragment--fragment interactions} \label{S: f-f inter}

	Fragments interact with each other in two distinct ways: they can merge or change orbits due to gravitational interaction and angular momentum transfer.

	Mergers (panel A1 of Fig. \ref{F: fragm-events}) occur when the fragments have a close encounter, usually in a collision orbit. As shown in the example provided in the panels A2, A3 and A4 of Fig. \ref{F: fragm-events}, fragments 42 and 40 are in a collision orbit and merge. The masses of the fragments combine, as shown by the spike in the mass of 40 in panel A4. We remark, however, several features of a typical merger. First, the mass of 42 has a small increase near the collision time. This can be easily explained by the fact that the mass integration domains start to overlap, since they are spheres of a fixed radius. The second remark is the mass decrease and small oscillations are observed after the collision. Collisions typically occur not head on, but with a certain impact parameter, which increases the spin angular momentum of the merged fragment. In the frame of reference of the fragment, the increased centrifugal force favors the development of secondary spiral arms, which can transport mass outwards, as is explained in Sect. \ref{S: subdisks}. During the fragmentation epoch in run x8, we recorded about 13 mergers in total. This kind of interactions can only be captured by simulating and resolving the full hydrodynamics of a fragment, and not by a sub-grid particle model.

	Changes in the orbit due to gravitational interaction of fragments that are unconnected by a spiral arm are somewhat more complex, less frequent and more difficult to determine. A general picture, however, is provided by panel B1 of Fig. \ref{F: fragm-events}: two fragments approach, and their gravitational interaction slows down fragment 1, therefore moving it to a lower orbit, and accelerates fragment 2, moving it to a higher orbit. Panel B4 provides an example of this kind of interaction: just after the orbits of 16 and 29 cross each other, 16 slows down and gets into a collision orbit with the central massive protostar, while 29 gets accelerated into a higher orbit. The moments leading to the orbit crossing, however, are another example of the effects of a gravitational interaction: in panels B2 and B3, we see how fragment 29, originally in a low orbit, migrates towards fragment 16 and vice versa, until they cross each other and the scenario illustrated as $t_0$ in panel B1 is reached. The example provided here comes from the data of run x4, but we also observed a similar behavior in run x8, but without the orbit crossing. Some of these interactions have also been observed previously in 2D simulations in the context of planet formation \citep[see, e.g.,][]{2012ApJ...746..110Z}. The outcome of these interactions (a change in the orbit or a merger) depends on the approach velocities and the mass ratio of the interacting fragments. In the example given in panels B2--4, the masses were comparable.

\subsubsection{Fragment interactions with spiral arms and the massive protostar} \label{S: accb}
	\begin{figure*}
		\centering
		\includegraphics[width=\textwidth]{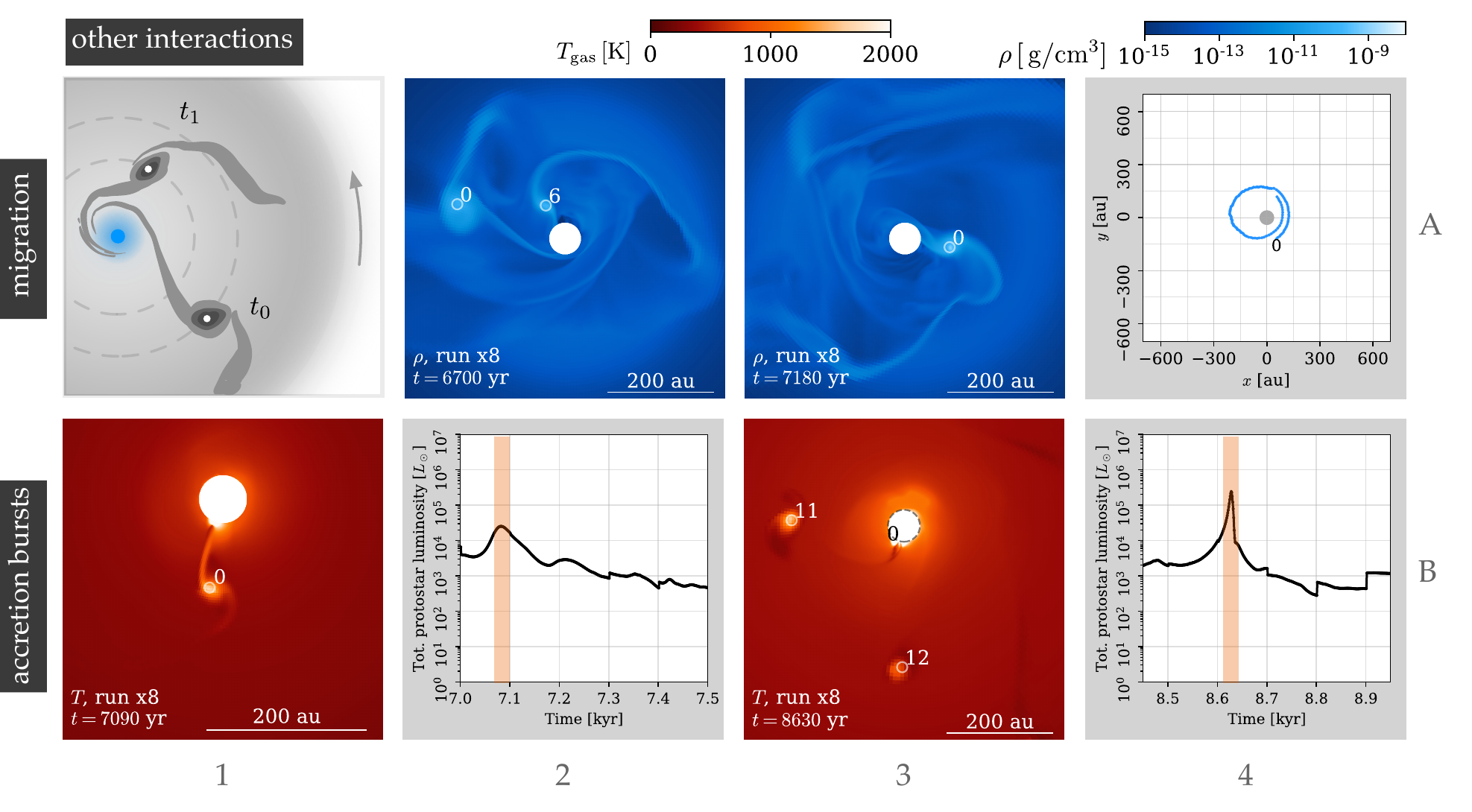}
		\caption{Interactions of the fragments with the spiral arms, and with the forming protostar modeled as a central sink cell. The orbit in panel A4 is plotted for a window of 2 kyr, with the fragment ID marking the starting point.}
		\label{F: fragm-events2}
	\end{figure*}

	The gravitational torques exerted by the spiral arms can cause migration, as shown by panels A1--4 of Fig. \ref{F: fragm-events2}. In this case, the eccentric orbit of fragment 0 reduces its periastron due to the action of the connecting spiral arm.
	
	Accretion onto the central massive protostar is not a smooth process, as depicted in the accretion rate shown in Fig. \ref{F: profiles}f. There are many discrete accretion events, caused by infall of matter through a spiral arm, or the complete accretion of a fragment. The transformation of gravitational energy into radiation is made in the form of accretion-driven bursts, sudden increases in the luminosity of the forming massive star (see Fig. \ref{F: starprop-x16}b). This phenomenon was reported theoretically in the context of massive star formation by \cite{2017MNRAS.464L..90M} and \cite{2019MNRAS.482.5459M}, where a system of magnitudes was also developed to describe them.

	Panels B1--4 of Fig. \ref{F: fragm-events2} show two examples of accretion bursts. Both the infall of matter through a spiral arm, as shown in panels B1--2, and the complete accretion of a fragment (panels B3--4), produce accretion bursts. The accretion of a fragment, however, produces a much sharper and brighter peak in luminosity compared to accretion through a spiral arm. Accretion bursts are also accompanied by overall increases in the temperature of the disk, that we term \textit{temperature flashes}, which typically last $\sim 30 \unit{yr}$ and heat up the fragments and spiral arms, as we discussed in the previous section.

	We must, however, discuss the effects of the inner boundary conditions (size of the central sink cell) on these accretion bursts. Since matter is only allowed to cross the sink cell inwards, we are not taking into account how outflows affect the accreted mass. Some fragments that the simulation shows as accreted into the sink cell could also only be in an elliptical orbit, as discussed in Sect. \ref{S: companion}, specially if they have already undergone second collapse, so they are less susceptible to shearing and subsequent accretion. These considerations could make strong luminosity accretion bursts less frequent and intense as presented.

\subsection{Secondary disks} \label{S: subdisks}
		\begin{figure}
		\subcaptionbox{}[0.48\columnwidth]{
			\includegraphics[width=0.48\columnwidth]{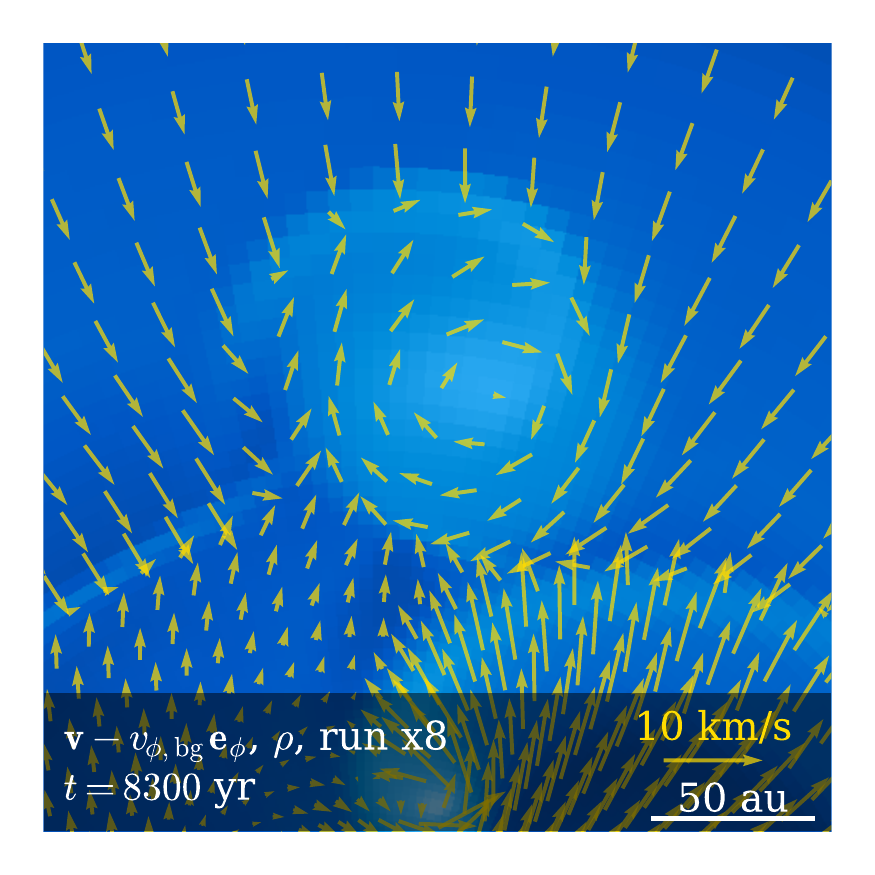}
		}
		\subcaptionbox{}[0.48\columnwidth]{
			\includegraphics[width=0.48\columnwidth]{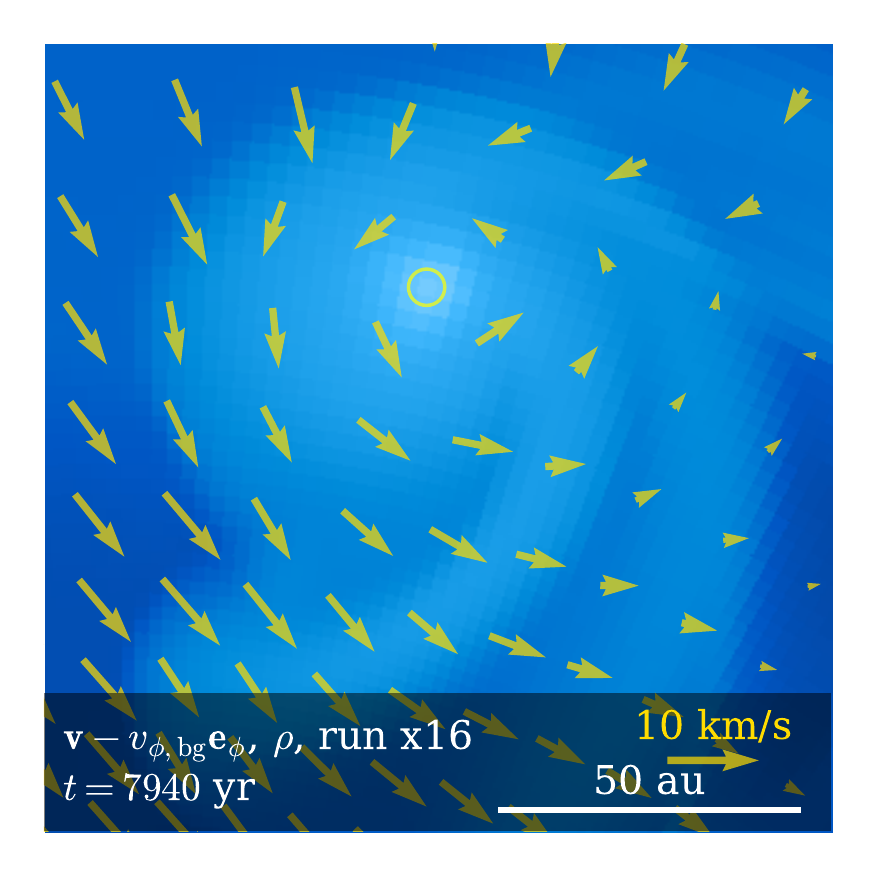}
		}

	\caption{Secondary disk with its spiral arms around (a) fragment 11 of run x8, and (b) fragment 12 of run x16. The background map represent density in the midplane (using the same color scale as the rest of the figures), and the arrows show the comoving velocity field. The central massive protostar is located beyond the bottom of the plotted areas in both cases. The yellow circle in (b) has a radius of $3\unit{au}$.}
	\label{F: subdisks}

	\end{figure}

	\begin{figure}
		\centering

		\subcaptionbox{Density, $t= 5.88 \unit{kyr}$}[0.95\columnwidth]{
			\includegraphics[width=\columnwidth]{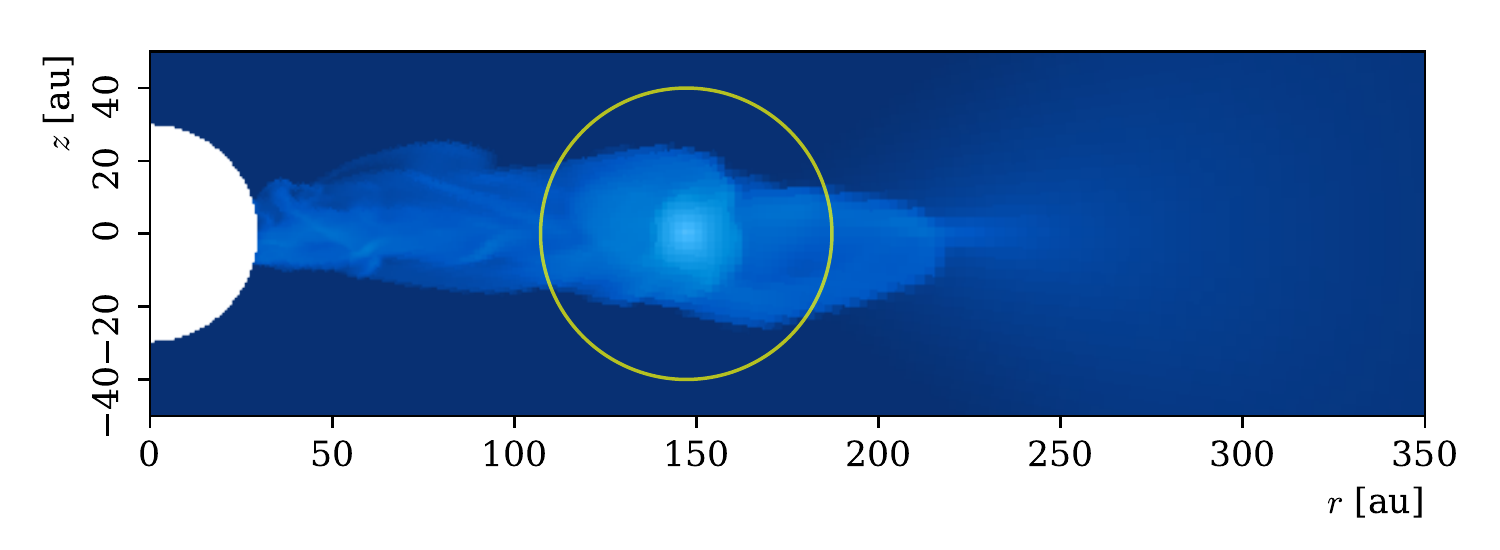}
		}
		\subcaptionbox{Density, $t = 7.12 \unit{kyr} $}[0.95\columnwidth]{
			\includegraphics[width=\columnwidth]{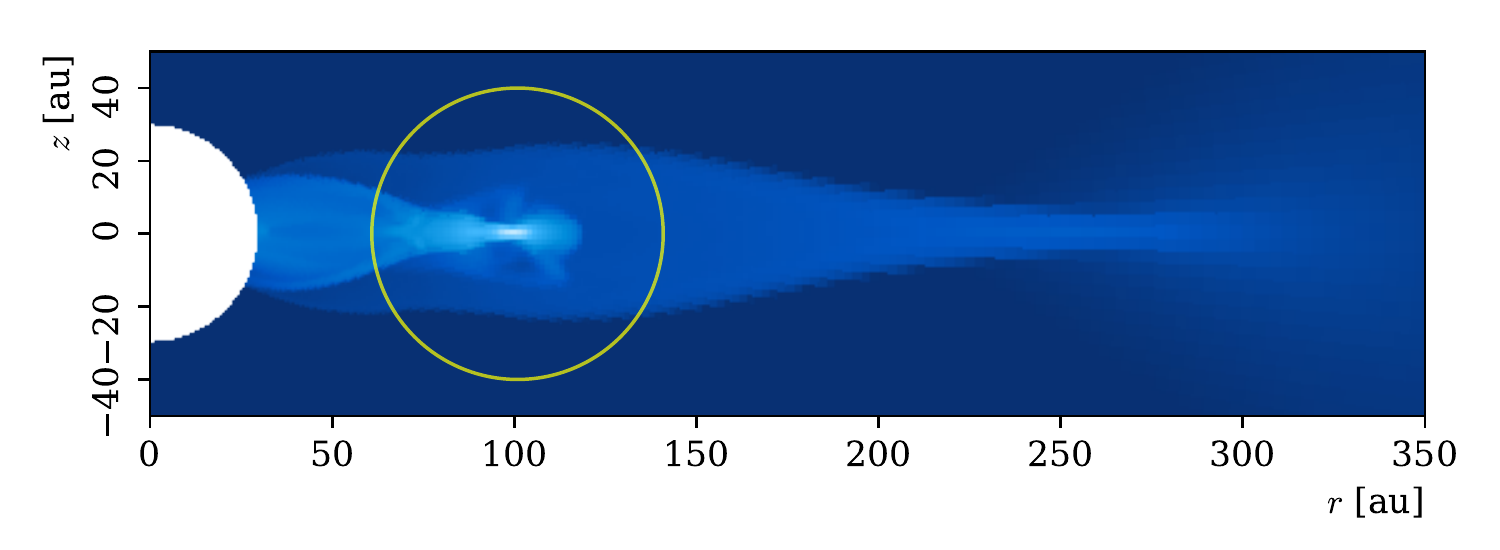}
		}

	\caption{Vertical structure of fragment 12 and its close surroundings (run x16). The yellow circle has a radius of $40\unit{au}$. The density uses the same color scale as the rest of the figures.}
	\label{F: vert-struct}

\end{figure}

	Fragments frequently develop a secondary disk with spiral arms, embedded in the primary disk, as shown in Fig. \ref{F: subdisks}. The figures show the comoving velocity field, $ \vec v - v_{\phi,\text{bg}} \vec e_\phi $, where $v_{\phi,\text{bg}}$ is the angular velocity profile of the primary disk. In Fig. \ref{F: subdisks}a, we observe a typical secondary disk, with the densest and hottest region (fragment) in its center. We also observe how converging flows form spiral arms. Converging flows form frequently when shearing and turbulent motion (due to activity of fragments and other spiral arms) produces a region with a net outward flow that encounters the inward radial flow from global mass transport.

	Spiral arms in secondary disks can also be drifted off the fragment, as shown in Fig. \ref{F: subdisks}b. The spiral arm encounters a decelerating inward flow that makes it drift apart. As a result, the calculated mass plot (Fig. \ref{F: fragm-mass-x8}) shows a decrease, since the mass of the spiral arm is lost from the fixed integration domain used. If the spiral arms are bounded to the fragment, the plotted mass of the fragment only shows an oscillation, but frequently the spiral arm gets absorbed by other spiral arms, provoking a real mass loss in the fragment, with a consequent temperature variation over time. This mechanism provides a way to transfer back some matter from fragments to the disk or the central massive protostar. This process can also trigger fragmentation, as was the case of the spiral arm that gives rise to fragment 59 shown in Fig. \ref{F: fragm-creat}A4.

	Figure \ref{F: vert-struct} shows the vertical structure of a fragment, using the highest resolution dataset. When the fragment is forming (Fig. \ref{F: vert-struct}a), it has a spherical-like shape. Fig. \ref{F: vert-struct}b, however, shows the vertical structure of the secondary disk shown in Fig. \ref{F: subdisks}a: the fragment indeed gets flattened down, with the highest density in the center.

\subsection{Comparison to observations}\label{S: comp-obs-x8}

\subsubsection{G11.92–0.61 MM 1}

Observations of the massive young stellar object G11.92–0.61 MM 1 reported in \cite{2018ApJ...869L..24I} show a fragmented Keplerian disk around a proto-O star. The main object MM1a, is reported to be $\sim 40 \unit{M_\odot}$ (with between 2.2 and 5.8$\unit{M_\odot}$ attributed to the disk, and the rest, to the protostar), and the fragment MM1b is reported to be $\sim 0.6 \unit{M_\odot}$, located at around $\sim 2000\unit{au}$ from MM1a.

This seems to be specially compatible to the general results in run x8, but later in time (the disk reported in the observations is bigger, and the central star is more massive). However, there are several warnings that should be taken into account. From the results in \cite{2018ApJ...869L..24I}, the mass ratio of the disk and the primary is between $0.055$ and $0.145$. In our simulations, the mass of the central massive protostar at $\sim 16\unit{kyr}$ is $\sim 20\unit{M_\odot}$ (cf. Fig.  \ref{F: starprop-x8}). We calculated the mass of the disk including the fragments and spiral arms (integration of density in a cylinder of $1500\unit{au}$ in radius and $20\unit{au}$ in height), and the mass of the background disk. The mass of the disk including substructures is comparable to the mass of the primary, as expected from a fragmenting disk; however, if substructures are excluded, the mass of the disk increases more slowly. After $\sim 16\unit{kyr}$, the mass of the background disk is $\sim 5 \unit{M_\odot}$, which means a mass ratio of $\sim 0.25$.  In addition, the surviving fragment (94) in run x8 is located at $\sim 1500\unit{au}$ from the primary, although at the end of the simulation it is moving outwards. The mass of fragment 94 is of the order of a few solar masses, although this value includes the mass of the secondary disk.

Given that the mass of most of the fragments produced in our simulations is of the order of $1\unit{M_\odot}$, the observed position and size of the disk, we think that the scenario described in \cite{2018ApJ...869L..24I} could plausibly be obtained with a setup similar to ours.

\subsubsection{Accretion burst event in G358.93-0.03}
Disk substructures associated to an accretion burst were observed and reported in \cite{2020NatAs.tmp..144C}, for the high-mass young stellar object G358.93-0.03. The flaring event was reported in \cite{2019ATel12446....1S}. \cite{2020NatAs.tmp..144C} performed a kinematic model that describes the accretion flow as occurring along two spiral arms, although they also acknowledge the compatibility of their observations with the accretion of a fragment. The results of our simulations support the second scenario, just as the results in \cite{2017MNRAS.464L..90M} and \cite{2018MNRAS.473.3615M} do. The processes described in Sects. \ref{S: fragm-orbits}, \ref{S: f-f inter} and \ref{S: accb}, namely, the gravitational interactions between fragments (as exemplified in panel B4 of Fig. \ref{F: fragm-events}), and the gravitational torques exerted by the spiral arms can cause the fragment to lose angular momentum, leading to  accretion by the primary, and thus causing an accretion burst.

\section{Companion formation} \label{S: companion}

As fragments contract by their self-gravity, their temperatures increase, which in turn causes an increase of their internal pressure, halting the collapse (classical Kelvin-Helmholtz contraction). Fragments whose central temperature goes beyond $\approx 2000 \unit{K}$, however, start to experience hydrogen dissociation, which means that gravitational energy is not converted into thermal kinetic energy, but rather it is used to dissociate hydrogen molecules; therefore allowing for further (\emph{second}) collapse, until a second Larson core is formed.

The exact temperature for hydrogen dissociation is density-dependent; we checked that for the central densities of the fragments, that temperature is between $1700\unit{K}$ (10\% dissociation) to $2300 \unit{K}$ (90\%) dissociation \cite[][]{2013ApJ...778...77D}. According to \cite{2018A&A...618A..95B}, second Larson cores formed from reservoirs of a few solar masses have radii of the order of a few solar radii. Based on the free fall timescale of a fragment of a few solar masses, we use the conservative estimate of $80 \unit{yr}$ for the duration of the second collapse. After the second core is formed, contraction continues to drive the temperature up, continuing the evolution of the protostar.

The simulations presented here do not include hydrogen dissociation. However, by tracking the central temperature of a fragment, one can determine if it should undergo second collapse and form a second core. This allows us to hypothesize the ultimate fate of the fragments if more realistic physics were considered.

\subsection{Central temperature of the fragments} \label{S: fragm-centraltemp}
	\begin{figure}
		\centering
		\includegraphics[width=\columnwidth]{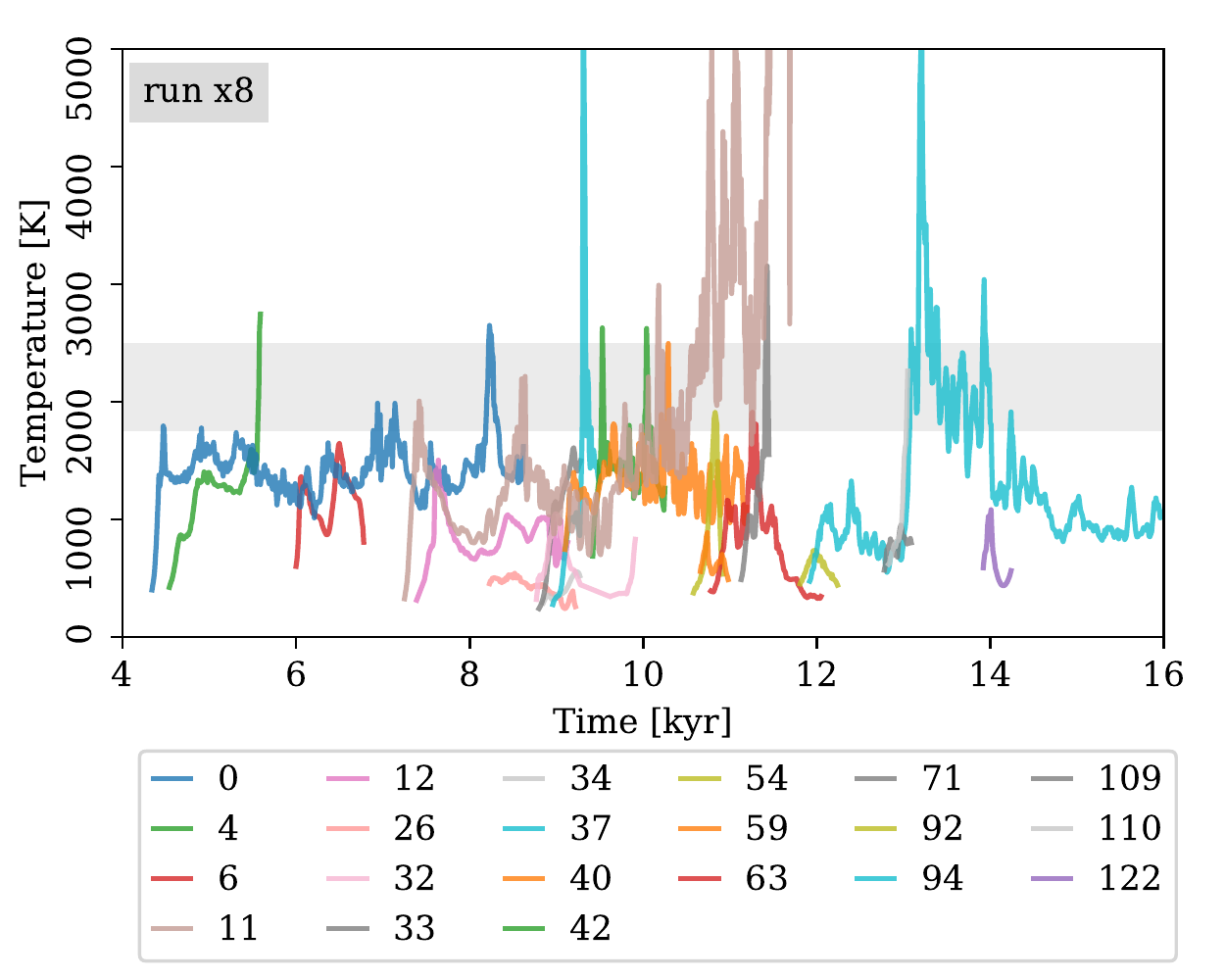}
		\caption{Temperature of the fragments with life span longer than 200 yr, for run x8. The gray box indicates the hydrogen dissociation limit.}
		\label{F: fragm-temp-x8}
	\end{figure}

		\begin{figure*}
		\centering
		\includegraphics[width=0.9\textwidth]{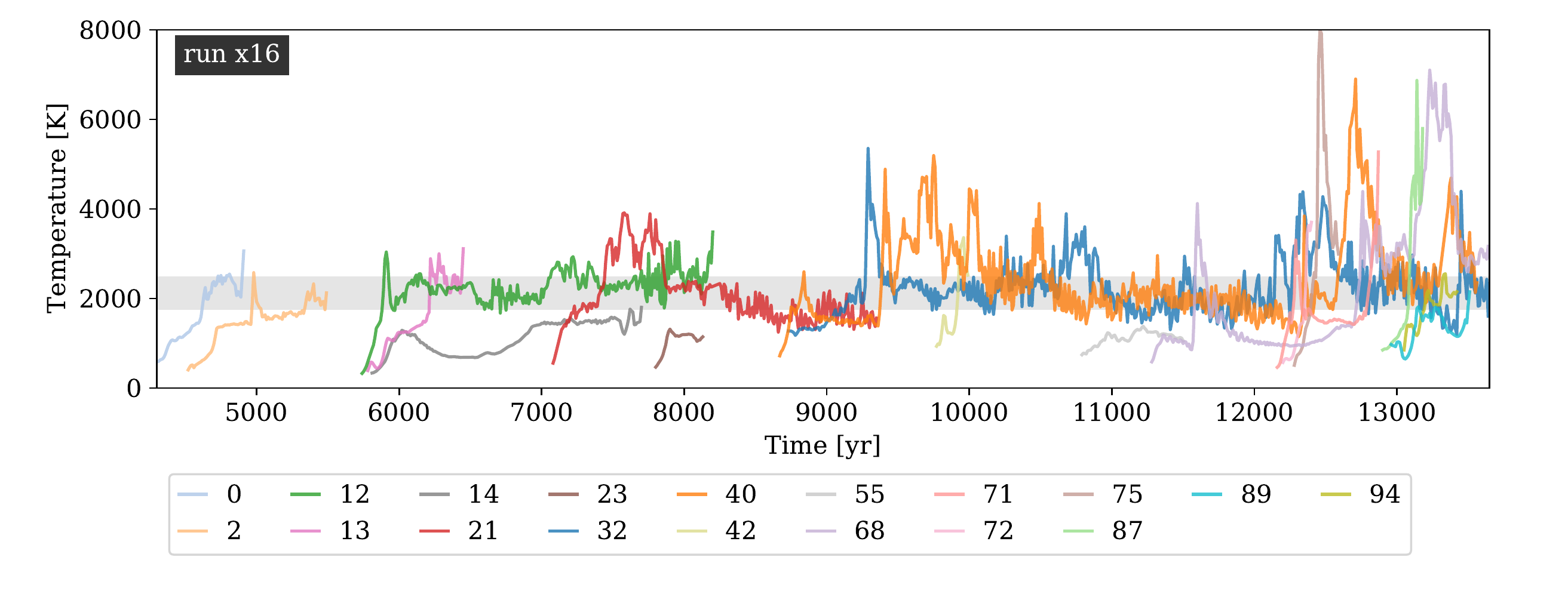}
		\caption{Temperature of the fragments with life span longer than 200 yr, for run x16. The gray box indicates the hydrogen dissociation limit.}
		\label{F: fragm-temp-x16}
	\end{figure*}

Figures \ref{F: fragm-temp-x8} and \ref{F: fragm-temp-x16} show the tracked central temperature of the fragments as a function of time, for runs x8 and x16, respectively. The shadowed box shows the threshold for hydrogen dissociation, which, as we mentioned, is not included in our simulations. This is why we see high temperatures, and fragments crossing the threshold both ways over time. In both figures, we see some high and short-lived spikes in the central temperature, that are a result of rapid mass increase, for example, due to mergers or accretion onto the fragment. The small periodic variations of the temperature are caused by the interactions between the fragments and their environment, including the development of secondary spiral arms. The mechanism of secondary spiral arm drift explained in Sect. \ref{S: subdisks}, that produces a lowering in the mass of the fragment, also produces a lowering in its temperature (due to decompression). The variation of the observed central temperature of the fragments with resolution is discussed in Sect. \ref{S: conv-fragm}.

Longer lived fragments tend, in general, to have higher temperatures of at least $\sim 1000 \unit{K}$. Fragments formed later in time tend to also reach higher central temperatures, since they have more mass (cf. Sect. \ref{S: fragm-mass}). 

	Fragments that cross the hydrogen dissociation temperature threshold undergo second collapse and become second cores. As a reference, we count the number of fragments that have a temperature higher than $\sim 2000\unit{K}$ for a time longer than $\sim 80\unit{yr}$, that is, our estimate of the time needed to form a second core. In run x8, 4 second core objects are formed, and in run x16, 10 second core objects are formed.

\subsection{Fate of the fragments} \label{S: fate}

Only after the second collapse has taken place, a fragment would reduce its radius from the order of a few au to a few stellar radii.  From the mere data of the simulations presented here, however, it is impossible to know if after the second collapse the temperature continues to rise and the second cores become actual companion stars. One point to take into account, however, is that many fragments develop secondary disks, which means that after the second collapse, more mass can be delivered to the second core and potentially allow a temperature increase and continuation of the evolutionary process.

Fragments of sizes of a few solar radii would be unresolvable with our current simulation setup. However, since their compactness would also provide stability, it would be safe to add a sub-grid particle model in order to study their behavior. Before the second collapse happens, however, fragments are just hydrostatic cores with strong interactions with the disk, spiral arms and other fragments. Some of these interactions are strong enough to destroy them. The next sections describe these mechanisms, and consider the effects that our simulation setup with a central sink cell have on the number of companions that are formed according to the simulations.

\subsubsection{Fragment destruction mechanisms}

\begin{figure}
	\centering
	\includegraphics[width=0.9\columnwidth]{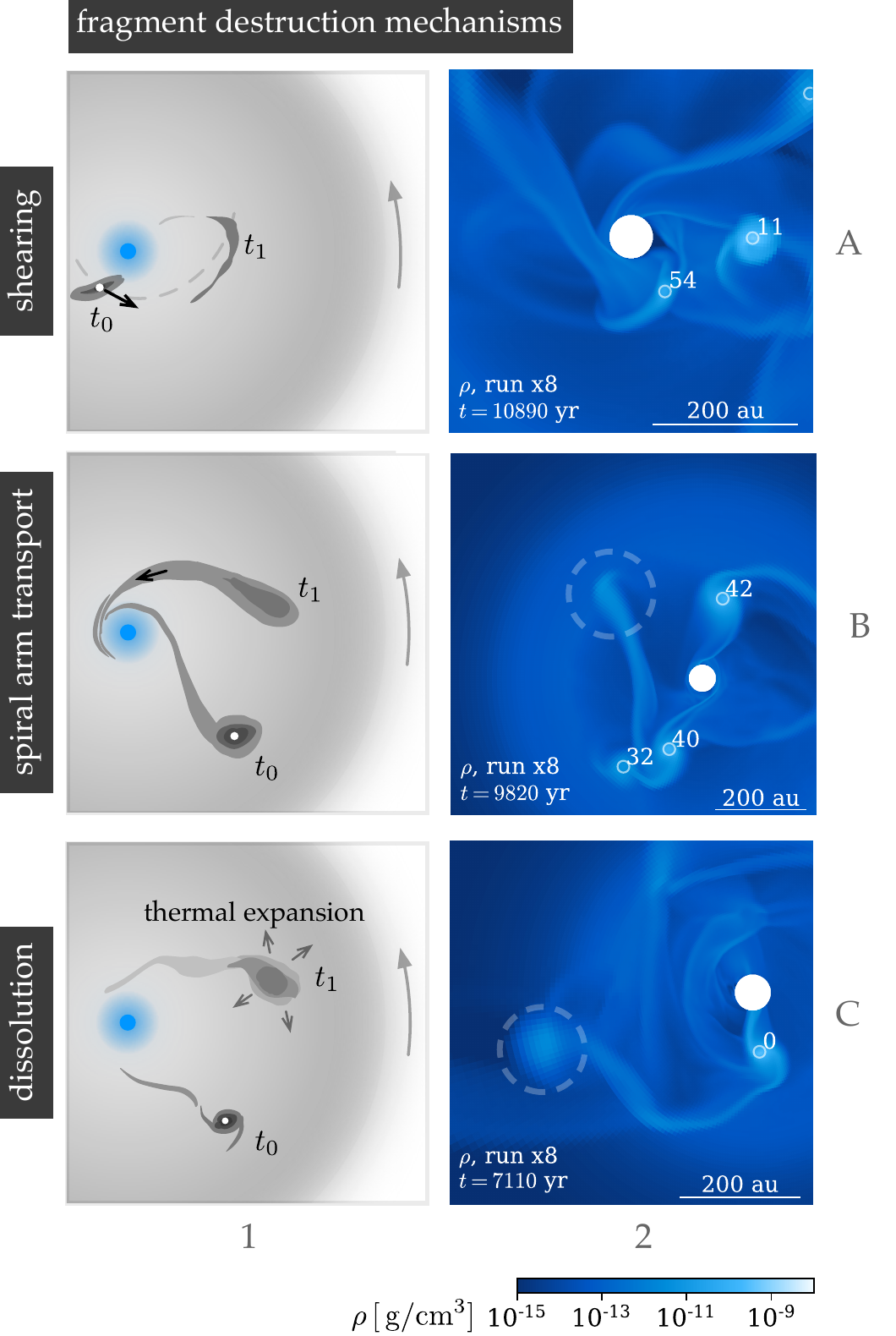}
	\caption{Fragment destruction mechanisms. }
	\label{F: fragm-destr}
\end{figure}

	Fragments that move in an eccentric orbit near the central primary protostar with high speeds experience shearing, which typically causes their destruction, as illustrated in panel A1 of Fig. \ref{F: fragm-destr}, and exemplified by panel A2. A similar process can be found in simulations on fragmentation in protoplanetary disks, as described in \cite{2015A&A...579A..32L}. It is also common in this scenario that some matter of the fragment gets accreted by the central protostar; sometimes forming a ``ring-like'' accretion structure like the one discussed in Sect. \ref{S: accretion disk}. In the particular example of panel A2, however, matter is not accreted. The sheared material expands and gets absorbed by spiral arms, providing converging flows needed for further fragmentation according to the mechanisms discussed previously.

	Spiral arms can also cause a mass loss to the fragment by transporting matter along them (Fig. B1 of Fig. \ref{F: fragm-destr}), eventually destroying the fragment. As an example, fragment 43 of run x8, shown in panel B2, is drained by a spiral arm that is being stretched, and delivers the material to fragments 32 and 40.

	Another fragment destruction mechanism occurs as a consequence of temperature flashes associated with accretion bursts (described in Sect. \ref{S: accb}). Fragments and spiral arms are heated up by the temperature flash. This triggers thermal expansion, which in turn lowers the density and central temperature of the fragment, compared with the values before the flash. If the fragment had an already low temperature and density, this thermal expansion and cooling can take the fragment below the detection threshold. After manually following the remaining matter, we observe it gets partially dissolved into the disk, absorbed into a spiral arm, absorbed by other fragments or it is sheared apart. An example of this behavior is provided by panels C1 and C2 of Fig. \ref{F: fragm-destr}.

	These destruction mechanisms severely affect the lifetime of fragments, if they are in their hydrostatic core phase, and show that sub-grid sink particle models may overestimate the number of formed companions. Only by resolving the sub-structure of the fragments these interactions become apparent.

\subsubsection{Fragment dynamics after the second collapse}

Once a second core is formed, the question arises on how will such an object interact with the environment, and, therefore, how the fragment destruction and merger mechanisms discussed above will impact the fate of a fragment. Although second cores would experience the same gravitational interactions than a hydrostatic core, they would not feel the same pressure gradient from the rest of the gas, due to their compactness. A second core would have less probability of colliding and merging with another second core, due to the reduction in the collisional cross section. A merger between a first core and a second core is a possible scenario: the first core would likely be sheared apart, and would form a denser secondary accretion disk around the second core, ultimately to be accreted by the latter.

\subsubsection{Formation of spectroscopic multiples}

\begin{figure}
		\centering
		\includegraphics[width=0.75\columnwidth]{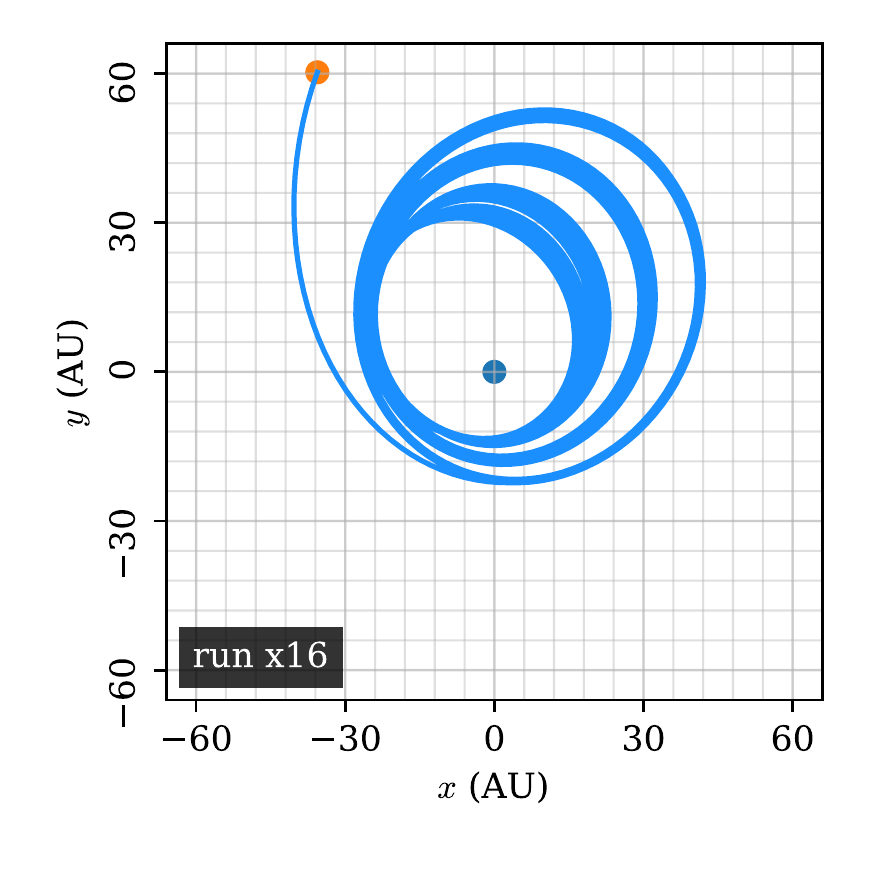}
		\caption{Integrated orbit of fragment 2 of run x16, after it enters the numerical sink cell. The orange dot indicates the initial position of the fragment, and the total time integrated was $3\unit{kyr}$. The sink cell (not shown) has a radius of $30\unit{au}$.}
		\label{F: final orbits}
\end{figure}

	As mentioned in Sect. \ref{S: pp-method}, a disadvantage of using a central sink cell is the inability to distinguish between accretion and the formation of close companions. This is specially true for fragments that have undergone second collapse, and therefore do not suffer the destruction mechanisms mentioned above. Once a fragment enters the numerical sink cell ($r=30\unit{au}$), its mass is counted as being accreted by the central massive protostar, when in reality, if the fragment is a second core, it might be in a orbit instead of a merger with the primary.

	Migration and gravitational interactions with other fragments are some of the responsible mechanisms for getting a fragment into an orbit closer to $30\unit{au}$ (the radius of the sink cell).  In order to study the possible fate of a second core, we integrated the orbit of fragment 2 of run x16 (Fig. \ref{F: final orbits}), which undergoes second collapse shortly before it enters the sink cell. In the integration, we considered the gravitational force of the central massive protostar and the gravitational force of the background disk, but no other interactions. Fragment 2 was originally dragged into the sink cell by gravitational torques arising from spiral arm interaction. The predicted integrated orbit is elliptical, and it gets smaller in time, due to the increase in mass of the central massive protostar. Some ``jumps'' can also be seen in the orbit, and they are caused every time the central massive protostar accretes (non-second-core) fragments, that is, its mass suddenly increases by a discrete amount. After $3\unit{kyr}$ of integration, the orbit of the fragment has a periastron of $\sim 15 \unit{au}$. At the time, the simulation is at $t=7.9 \unit{kyr}$, and the mass of the central protostar is $\sim 4\unit{M_\odot}$. As time progresses and the central protostar becomes more massive, the orbit of the fragment should become smaller and smaller, thus providing a mechanism for the formation of spectroscopic multiples.

\subsection{Implications}

\subsubsection{Accretion bursts} \label{S: impl-accb}

\begin{figure}
	\subcaptionbox{}[0.48\columnwidth]{
			\includegraphics[width=0.48\columnwidth]{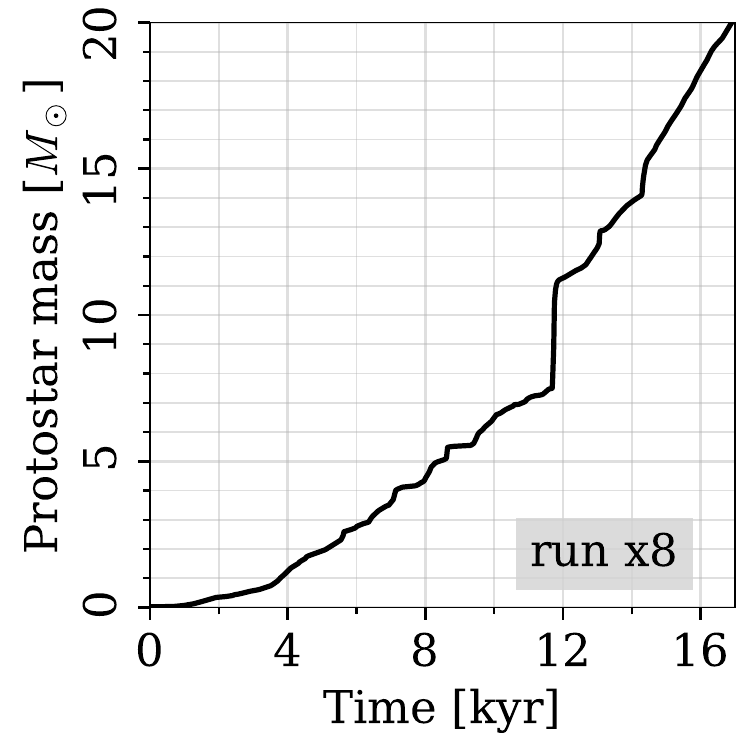}
		}
	\subcaptionbox{}[0.48\columnwidth]{
			\includegraphics[width=0.48\columnwidth]{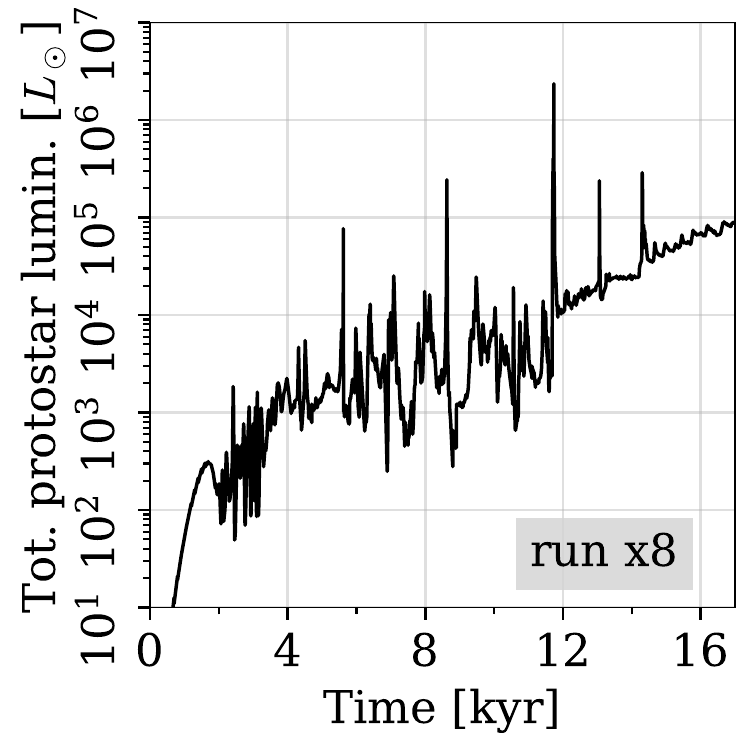}
		}
		\caption{(a) Mass and (b) luminosity of the central massive protostar, as formed in run x8.}
		\label{F: starprop-x8}
\end{figure}

	In run x8, fragment 11 reaches the hydrogen dissociation temperature at $10.22\unit{kyr}$ and would undergo second collapse if the simulation had implemented it. At the time, the mass of the fragment is $1.8\unit{M_\odot}$. Then, at $11.47\unit{kyr}$, it merges with fragment 71, which greatly increases the temperature (cf. Fig. \ref{F: fragm-temp-x8}), and its mass briefly reaches $5\unit{M_\odot}$. This would be an example of a second core merging with a hydrostatic core, and we would expect, in a more realistic setup, that fragment 11 would at least partially accrete the material from fragment 71 through the secondary accretion disk. The interaction between fragments 11 and 71, however, alters the orbit of fragment 11 and it ultimately reaches the sink cell. A second core entering the sink cell might produce, however, a companion instead of an accretion event. For the plots of run x8 presented in this paper, this means that, at $t\sim 11.5\unit{kyr}$:
\begin{itemize}
\item the high peak in the accretion rate (Fig. \ref{F: profiles}f), the $\gtrsim 10^6 \unit{L_\odot}$ accretion burst shown in Fig. \ref{F: starprop-x8}b and the bump in the mass of the central massive protostar (Fig. \ref{F: starprop-x8}a) might not take place;
\item the subsequent inner disk overdensity that arises in the radial profiles for run x8 (Figs. \ref{F: profiles}a, \ref{F: profiles}b and \ref{F: profiles}e) during the interval $[12,16[\unit{kyr}$ as a result the shearing and accretion of fragment 11 might not be there.
\end{itemize}

	According to the preceding discussion, very high luminosity bursts are less probable to occur than what is shown in Fig. \ref{F: starprop-x16}b, since they require the accretion of a significantly large mass in a short time. Fragments that have the required mass have most likely undergone second collapse, and, according to the integrated orbits we presented above, they will likely form close companions instead of being accreted. The possibility that a second core be accreted is not excluded, but it should be rare due to collisional cross section considerations.

\subsubsection{Number of companions}

	Taking into account the preceding elements, we summarize here the results of our simulations regarding companion formation.

	There are 4 second core objects produced in run x8. From them, one (fragment 94) survives in the outer disk ($r \sim 1400\unit{au}$ at the end of the simulation). Fragments 0 and 11 form close companions, and fragment 37 merges with another fragment (so its fate is unclear from our simulations).

	In run x16, there are 10 second core objects produced. Fragments 0, 12 and 13 form close companions, and fragments 40, 32 and 68 survive the simulation in the middle region of the disk (see discussion in Sect. \ref{S: fragm-orbits}). The rest merge with other fragments, so, again, their fate is uncertain from our simulation.

	By analyzing which of the fragments that get close to the central massive protostar will produce accretion bursts and which ones satisfy the conditions for further stellar evolution, we have found that the number of possible close companions produced in our simulations is low. This result is consistent with observations of more evolved systems, such as \cite{2012Sci...337..444S}, \cite{2014ApJS..213...34K} and \cite{2015A&A...580A..93D}, where spectroscopic binaries or multiples are present in large fractions, but they also do not observe tens or hundreds of spectroscopic companions to a central massive star. Our results show, however, that companions produced by disk fragmentation can also exist at distances of the order of $\sim 1000\unit{au}$ astronomical units (although inward and outward migration is still possible in the long term). We also note in this comparison that the masses of the fragments given in the preceding sections are not necessarily an indication of the masses of the future companions, since accretion is still ongoing at the end of our simulations.

	% - - - - - - - - - - - - - - - - - - - - - - - - - - - - - - - - - - - - - - - -
\section{Resolution effects} \label{S: numerical}
	% Convergence of number of fragments, temperature of the fragments

	\begin{figure*}
		\includegraphics[width=0.95\textwidth]{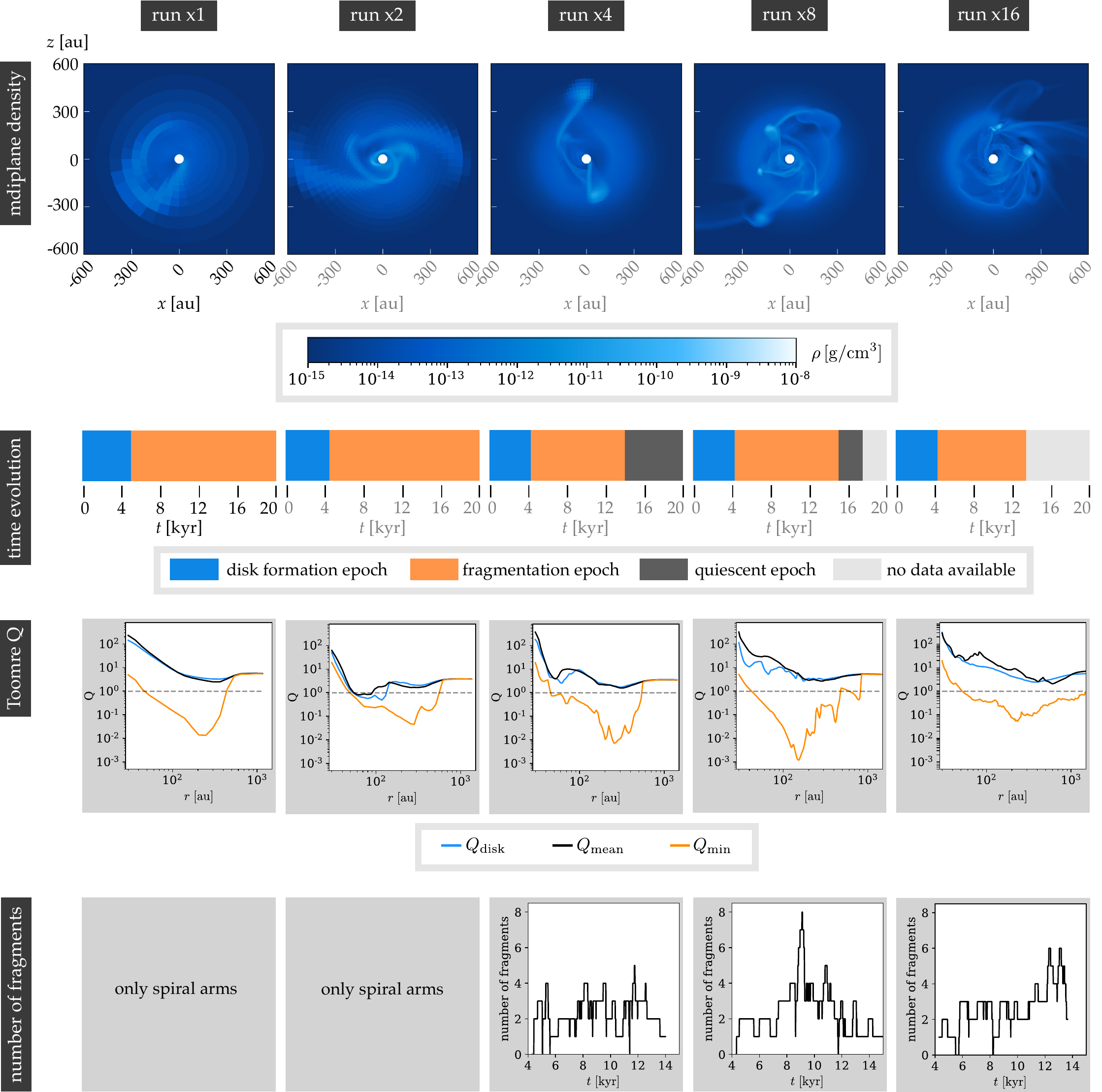}
		\caption{Convergence of different quantities with resolution. The midplane density data was taken at $t=7\,500\unit{yr}$ for all the maps in the row. The Toomre Q was time-averaged during the fragmentation epoch.}
		\label{F: fragm-convergence-1}
	\end{figure*}

	\begin{figure}
		\centering
		\includegraphics[width=\columnwidth]{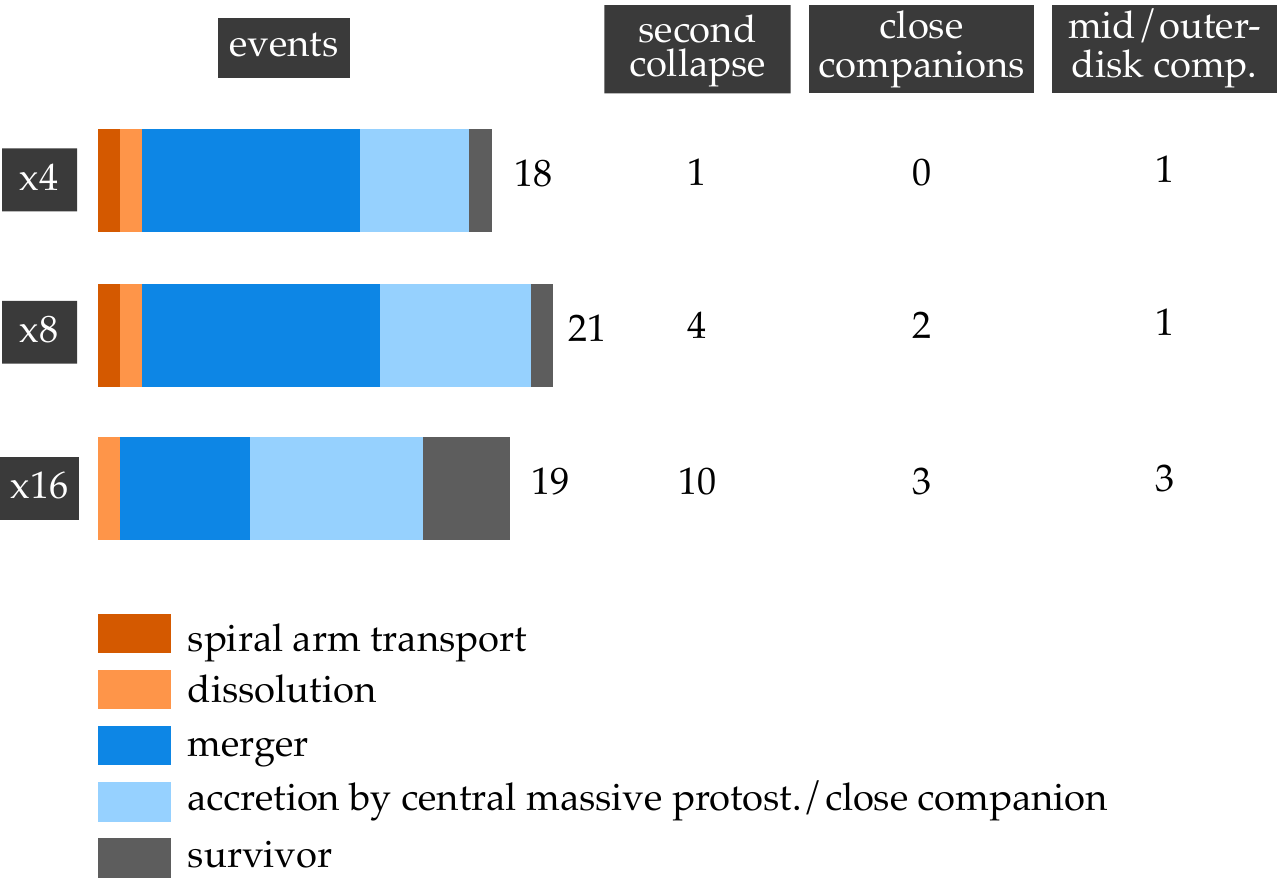}
		\caption{Fragment statistics for the fragment-producing runs. The charts show the fate of the fragments with a life longer than 200 yr according to the simulation, that is, without taking into account the effects of the inner boundary and hydrogen dissociation, as discussed in Sect. \ref{S: fate}. The estimated number of these fragments that undergo second collapse are given on the right column.}
		\label{F: fragment-convergence-2}
	\end{figure}

	After describing the main results, the focus of the next section is in describing how these change with resolution by comparing five runs with different spatial resolution.

\subsection{Convergence of fragmentation}
	During this subsection, we will refer to the panels in Fig. \ref{F: fragm-convergence-1}. The midplane density maps (Fig. \ref{F: fragm-convergence-1}) show that only spiral arms are produced in runs x1 and x2, but no fragments. By no fragments we mean that our conditions for fragment detection are not met, and that the shape of the substructures formed (regions of higher density than the background disk) is difficult to be recognized as a fragment because of resolution. Runs x4, x8 and x16 produce fragments and spiral arms. Runs x1, x2 and x4 show thicker spiral arms than runs x8 and x16, probably as a result of numerical diffusion. Substructure formation in runs x1 and x2 continues until the end of the simulated time, that is, the evolution of the system does not drastically change over time. In contrast, a quiescent epoch with no further fragmentation is observed in runs x4 and x8. Due to computational costs, data for run x16 is not available beyond $t=13.5\unit{kyr}$, not enough to confirm the existence of a quiescent epoch in that run as well. In summary, only the runs with high enough spatial resolution exhibit an end to the fragmentation epoch.

	The Toomre $Q$ parameter shows a similar behavior for all resolutions. We had seen in Fig. \ref{F: Q_local_global} that the value of $Q$ for fragments and spiral arms, corresponding to the $Q_\text{min}$, shows the susceptibility of these areas to fragmentation, but the background disk and mean values (expected to be dominant in underresolved observations) show the system as Toomre-stable.

	The number of fragments that live longer than 200 yr present on a given time in the simulation is also similar, except for the peak of 8 fragments seen in simulation x8, and that we discussed earlier. The average number of fragments at a given time during the fragmentation epoch is around $2.2$. The tendency of longer-lived fragments to be created at larger radial positions with time is also observed in the other simulations, although not shown here.

\subsection{Convergence in the properties of the fragments} \label{S: conv-fragm}

	The fate of the fragments without taking into account the effects of the inner boundary and hydrogen dissociation we discussed in Sect. \ref{S: fate} are shown in Fig. \ref{F: fragment-convergence-2}. Mergers and shear/accretion into the sink cell dominate the mechanisms of fragment destruction. Spiral arm matter transport and dissolutions are far more rare.
	
	Comparing Figs. \ref{F: fragm-temp-x8} and \ref{F: fragm-temp-x16}, the central temperature of fragments gets higher with resolution. From this, the number of fragments that get to the second collapse phase increases from run x8 to run 16, but it remains in the same order of magnitude.  In \cite{2018A&A...618A..95B} (Fig. 2c), the temperature profile of a first core shows that the maximum temperature is reached at a radius of $\sim 1\unit{au}$. Run x16 has a resolution of $\lesssim 1\unit{au}$ at $r \lesssim 100\unit{au}$, which might indicate that the values of the central temperature for fragments in the inner disk are close to convergence. 
	
The statistics of fragment interactions show that independently of resolution, fragments in a hydrostatic core stage are fragile, and it is not safe to include them in a particle sub-grid model if they do not reach second core status, for which the correct resolution is needed. As already discussed in Sect. \ref{S: fragm-mass}, the masses of the fragments for runs x8 and x16 are similar, and although it is not shown here, the masses of the fragments produced in run x4 also have consistent values.
		
	Fragments in the outer disk, as exemplified by fragment 94 of run x8, tend to have higher masses, and therefore, higher probabilities of reaching the temperature required for second collapse. However, the spatial resolution of the numerical grid is lower in the outer disk (for run x8, the cell size is $\sim 20\unit{au}$ at $\sim 1000\unit{au}$); and although we show in Sect. \ref{S: resolution} that, at that radial distance, this is enough to resolve the Jeans length and therefore the formation of fragments, it is not enough to resolve the substructure of the fragments. Therefore, fragments located at $r\gtrsim 1000 \unit{au}$ have radii that are artificially higher due to numerical diffusion, even when not considering the physics of the second collapse. A larger radius makes the fragment less gravitationally bound, and filaments and secondary spiral arms are developed between the fragment and the central massive protostar. These structures fragment, the fragments migrate rapidly inwards and produce accretion bursts in the late stages of runs x4 and x8, during the quiescent epoch, but we believe this effect to be mainly numerical. Due to better spatial resolution, fragments produced in run x16 have more consistent sizes across the disk.

\subsection{Convergence of the formed massive protostar}

	\begin{figure}
				\centering
		\subcaptionbox{}[0.45\textwidth]{
			\includegraphics[width=0.45\textwidth]{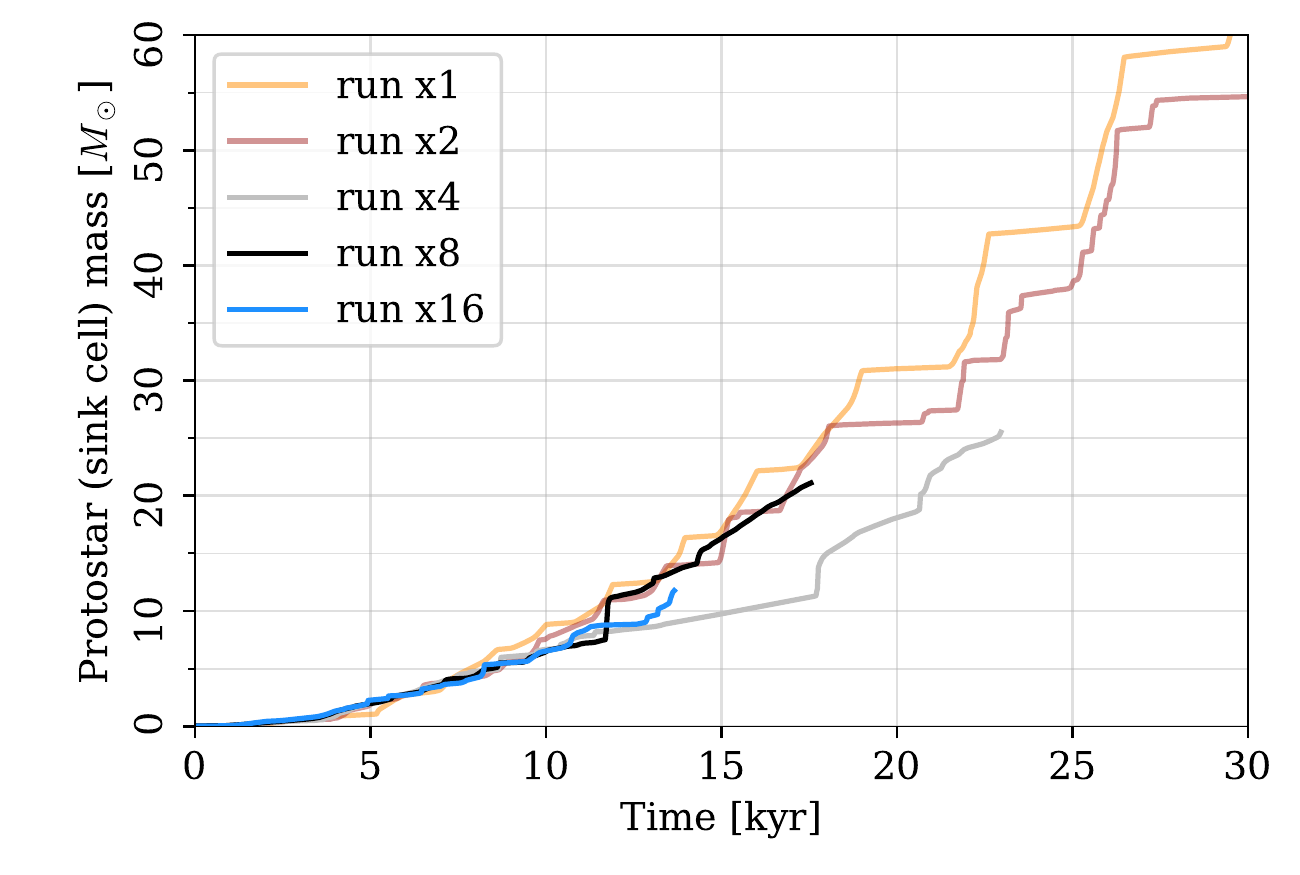}
		}
		\subcaptionbox{}[0.45\textwidth]{
			\includegraphics[width=0.45\textwidth]{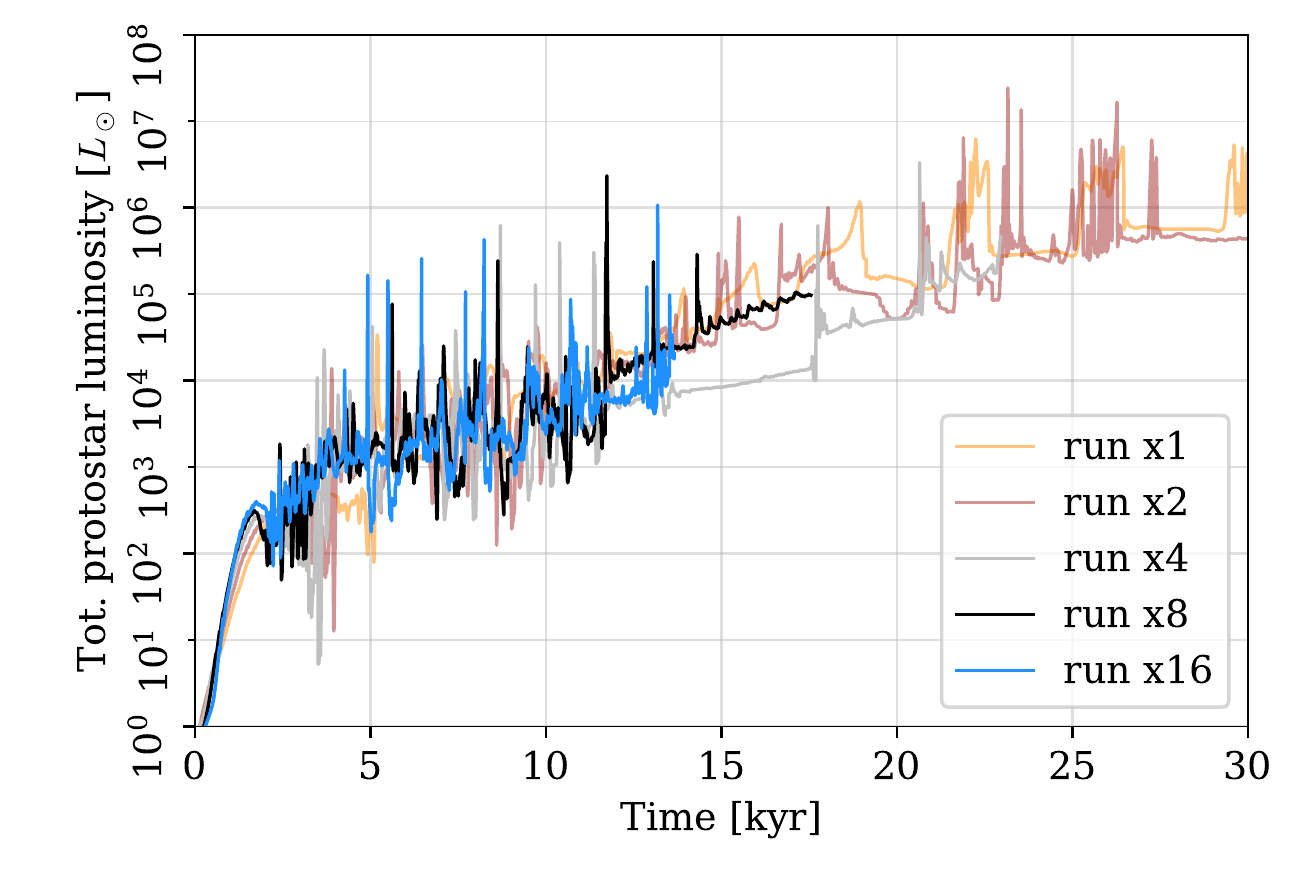}
		}
		\caption{Convergence of the properties of the formed protostar.}
		\label{F: converg-protostar}
	\end{figure}

	During the simulation, several properties of the central massive protostar are calculated, under several assumptions. Figure \ref{F: converg-protostar} shows the mass and total luminosity. We remind the reader that the mass of the central protostar is simply the mass of the sink cell, and the total stellar luminosity is the sum of the accretion luminosity (total conversion of gravitational potential energy of the accreted material into radiation) and the stellar luminosity calculated following the evolutionary tracks of \cite{2009ApJ...691..823H} (that depend on the mass of the protostar and the accretion rate).

	Figure \ref{F: converg-protostar}a shows only qualitative convergence in the protostellar mass after $\sim 11\unit{kyr}$, since it depends heavily on how the material is accreted in each simulation, and this, in turn depends on the fragmentation that is developed in time. Figure \ref{F: converg-protostar}b, shows that the total luminosity curves are similar for all simulations. All feature some accretion bursts and a general increase in time due to the contribution of stellar evolution. The plots for run x4 and x8 also show a general reduction in accretion bursts at the end of the simulated time (quiescent epoch), except for some bursts caused by the effect discussed in Sect. \ref{S: conv-fragm}, triggered by fragments in the outer disk. There is no quiescence observed for runs x1 and x2. We also find that the central massive protostar starts burning hydrogen at around $12\unit{kyr}$ for runs x16 and x8, $\sim 18\unit{kyr}$ for run x4, and $\sim 16\unit{kyr}$ for runs x1 and x2.

\section{Comparison with previous studies} \label{S: literature}

\subsection{Overview}

	Several studies on massive star formation, including cloud and disk fragmentation, have been performed over the years. We summarize some of them in this section, and we will compare their outcomes and main features in the following subsections.
	
	First, we discuss some studies on cloud fragmentation.  \cite{2012MNRAS.420..613G} performed several gravito-hydrodynamical simulations to study a collapsing cloud under the fragmentation-induced starvation scenario without radiation transport, an AMR grid of a minimum size of 13 au, and sink particles. In these simulations, various initial density profiles were used, together with a supersonically turbulent initial velocity profile but no rotation. They find between 161 and 429 sink particles in total; some of their simulations show filamentary accretion, and some, the formation of an accretion disk.
	
	  \cite{2010ApJ...725..134P} and \cite{2011ApJ...729...72P} used the code FLASH to study the effects of magnetic fields and ionizing radiation in a large-scale collapsing cloud of $ 1000 \unit{M_\odot} $, but neglecting the thermodynamics of dust re-emission, i.e., continuum radiation transport is not taken into account. Both studies used AMR with a minimum cell size of 98 au, and sink particles. According to our results, however, these sink particles could not be comparable to our fragments, since their accretion radius of $\sim 400 \unit{au}$ is at least one order of magnitude bigger than the size of a fragment with the potential of forming a companion, and more comparable to the size of the whole primary accretion disk, that is $\sim 400\unit{au}$ in radius in the middle of the fragmentation epoch. With no magnetic field, around 25 sink particles were created; the presence of magnetic field reduced fragmentation by around a factor of 2, and generated a more massive protostar, due to magnetic braking. Similar results were obtained by \cite{2011A&A...528A..72H}, where the RAMSES code was used to study a cloud of initial mass $ 100 \unit{M_\odot} $ and radius of $1.35\unit{pc}$. They used AMR with a minimum cell size of 8 au and 2 au for their low and high resolution runs, respectively, and no sink particles, but use the barotropic equation of state. The number of fragments obtained is of the order of 50 without magnetic field.

	The following studies considered also disk formation and fragmentation. In \cite{2007ApJ...656..959K} and \cite{2009Sci...323..754K}, three-dimensional gravito-hydrodynamical simulations were performed, with a gray flux-limited diffusion approximation for stellar radiation as well as radiation by the dust. The first paper studied the collapse of a core in initial supersonic shock-dominated turbulence, while the second paper used solid body rotation, without any initial turbulence. Both studies used sink particles following the Jeans criterion only. In the turbulent case for an initial mass of $200 \unit{M_\odot}$ \citep{2007ApJ...656..959K}, the spherical cloud forms filaments that feed two clumps that become in the end one massive protostar with an accretion disk that fragments in spiral arms, but no further fragmentation; 3 sink particles, however, are independently formed at around 3000 au from the primary. With regards to fragmentation, in \cite{2009Sci...323..754K}, the disk forms earlier, and at the end, the system develops into a binary of separation $\sim 1500\unit{au}$. Both simulations used the code Orion, with a AMR grid of maximum resolution of 10 au.

	\cite{2016ApJ...823...28K} used the FLASH code with an improved treatment of radiation transport, and an AMR grid with minimum cell size of 10 au. The cloud was initially in solid body rotation, and masses of 30, 100 and $ 200\unit{M_\odot} $ were considered. They used a stricter sink particle algorithm that, in addition to the Jeans criterion, checks for a convergent flow, a gravitational potential minimum and a negative total energy. The use of this criterion with the resolution considered lead to no fragmentation at all; only the formation of spiral arms was observed.

	\cite{2016MNRAS.463.2553R} repeated the simulations for both the initially laminar and highly supersonic turbulent cases, including a hybrid radiation transfer method that properly treats the multi-frequency stellar irradiation and gray flux-limited diffusion thermal re-emission \citep{2017JCoPh.330..924R}, but the maximum resolution of their AMR grid was 20 au. Their sink particle algorithm detects 29 companions, 16 with masses $ > 0.1 \unit{M_\odot} $ for the laminar case, and 3 companions for the turbulent case. In both cases, an accretion disk with spiral arms is formed around the massive protostar, but in the turbulent case, the disk becomes eccentric. Contrary to \cite{2009Sci...323..754K}, the system does not form a binary (in part due to different sink particle merging criteria). Initially, in the turbulent case, the massive protostar is fed primarily by filaments, however, the accretion rate is not significantly different than in the laminar case. A similar setup was used by \cite{2019ApJ...887..108R} to study the role of turbulence in fragmenting cores with initially virial and subvirial initial conditions, finding that virialized cores undergo significant turbulent fragmentation at early times, compared to subvirial cores. In both cases, a fragmenting accretion disk was formed.

	\cite{2017MNRAS.464L..90M, 2018MNRAS.473.3615M} used Pluto with a setup similar to ours, i.e., a fixed spherical grid with the radial coordinate increasing logarithmically. They also used the same radiation transport scheme, but lower spatial resolution. In \cite{2018MNRAS.473.3615M}, several initial angular velocity profiles were examined for fragmentation, specifically, $\beta_\Omega = 0$, $-0.35$ and $-0.75$. They found formation of spiral arms, followed by fragmentation; the highest number of fragments was obtained for $\beta_\Omega = -0.75$. It was proposed that fragmentation might explain the high spectroscopic binary fraction of massive stars. Accretion bursts were observed in \cite{2017MNRAS.464L..90M, 2018MNRAS.473.3615M, 2019MNRAS.482.5459M}. \cite{2019MNRAS.482.5459M} also studied the effects of disk wobbling during the fragmentation epoch.

\subsection{Resolution of the grid and the use of sink particles} \label{S: resolution}

		\begin{figure}
		\includegraphics[width=\columnwidth]{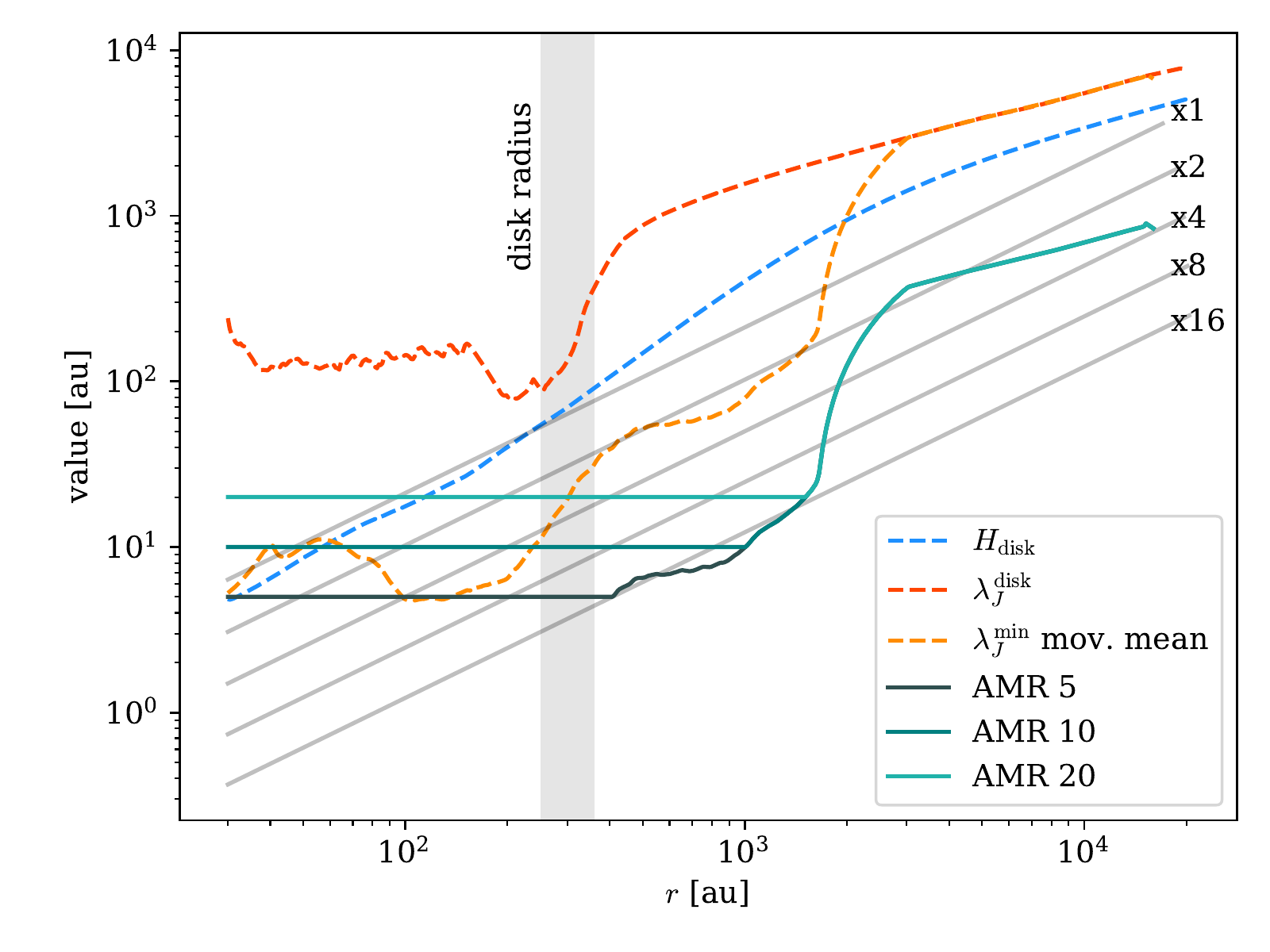}
		\caption{Comparison of midplane grid cell sizes in our fixed grid and other studies that use AMR. The different relevant scales of the system are shown in dashed lines; they were calculated as averages for the time period [6,8] kyr, and the corresponding disk radius is shown in the gray box. The number following the label for the AMR curves indicates the minimum cell size in au.}
		\label{F: resolution-comparison}
	\end{figure}

In order to check whether a simulation is resolving the correct scales associated with the physical phenomena described, we computed the pressure scale height of the disk and the Jeans length using the data from run x16, averaged over the time period $[6,8]\unit{kyr}$, which corresponds to the fragmentation epoch. The pressure scale height gives an idea of the scale of the vertical structure of the disk, and it is defined in the midplane as

\begin{equation}
\frac{H_\text{disk}}{r} = \frac{c_s}{v_\phi}
\end{equation}

where $c_s^2 = \gamma \, \partial P / \partial \rho$ is the sound speed. The Jeans length $\lambda_J$ indicates the size of a region that becomes gravitationally unstable at a certain density $\rho$ and temperature, and it is calculated as

\begin{equation} \lambda_J = c_s \sqrt{\frac{\pi}{G\rho}} . \end{equation}

In order to adequately study fragmentation numerically, the Jeans length has to be resolved properly. In order to study a Jeans length radial profile, we separate it into two values: the Jeans length of the background disk $\lambda_J^\text{disk}$, calculated with the median values (as described in Sect. \ref{S: pp-method}), and the minimum Jeans length, $\lambda_J^\text{min}$, that corresponds to the value of the most fragmentation-prone areas of the disk, i.e., the fragments and spiral arms.

Figure \ref{F: resolution-comparison} contains the radial profiles of these quantities, calculated using the data from run x16, and time-averaged during the period between 6 and 8\,kyr. The disk radius is also shown as a reference. Additionally, the cell size of our simulations, $\Delta x(r)$, is shown for each run (gray lines).

The pressure scale height scales approximately linearly with distance, giving an ansatz on the required cell size needed to resolve the vertical structure of the disk. The cell size of our logarithmic grid also scales linearly with distance. As a result of the simulation, we obtain that, on the disk region, $H(r) \approx 0.12 r^{1.07}$, and a comparison with the different runs yields that $H$ is resolved by all of them, except partially by x1.

The Jeans length decreases with increasing density, which means that it increases with distance. Our choice of the coordinate grid, then, makes possible a better resolution of the Jeans length while saving computational power. The Jeans length for the background disk is resolved by all runs, and indeed, we observe some sort of structure formation (spiral arms and/or fragments) in all runs. However, when observing the curve for the minimum value of the Jeans length, it is clear that only runs x8 and x16 are able to resolve fragmentation adequately, and run x4 only resolves fragmentation partially, which coincides with the results found in the convergence study presented in Sect. \ref{S: numerical}.

In order to compare the results from other studies, we establish approximate equivalences between the different grids used. In the case of the fixed grids used in \cite{2017MNRAS.464L..90M, 2018MNRAS.473.3615M, 2019MNRAS.482.5459M}, it goes as follows: (a) \cite{2018MNRAS.473.3615M} use approximately the same grid structure of our run x4 (except in their 'Run 1-HR', where the grid used is almost our run x8); (b) \cite{2017MNRAS.464L..90M} and \cite{2019MNRAS.482.5459M} use a grid comparable with our run x4.

A direct comparison of our fixed grid with AMR implementations is not trivial, since they are dependent on the selected refinement criteria and the particular evolution of the system. In order to give an idea of the resolution differences, we take the criteria given in, e.g., \cite{2016MNRAS.463.2553R}, i. e., a Jeans length should be resolved by at least 8 cells. The resulting profiles are shown in green in Fig. \ref{F: resolution-comparison}, where if 8 Jeans lengths cannot be resolved, the maximum refinement is plotted.

Now we examine the disk region. According to our results, the minimum Jeans length is only marginally resolved by an AMR simulation that has a minimum cell size of 5 au. Even in that case, our simulations provide much more detail, especially in the region $100\unit{au} \lesssim r \lesssim 600$, where according to Fig. \ref{F: num-fragm}b we observe the biggest number of long-lived fragments being created. This choice has allowed us, as we have shown in the results, to study even the interactions between fragments and their internal structure, and not merely to resolve where a fragment should be created by the Jeans criterion.

	Based on these considerations, we can establish that an AMR grid with a minimum cell size of $5\unit{au}$ provides a resolution between runs x16 and x8 for the outer disk, while a resolution similar or lower than run x4 for the inner disk ($r\lesssim 100\unit{au}$). A minimum cell size of $10\unit{au}$ corresponds to a resolution between runs x8 and x4 for the outer disk, but to run x2 for the inner disk. A minimum cell size of $20\unit{au}$ provides a resolution between run x4 and run x2 for the outer disk, but a resolution similar or lower than run x1 for the inner disk.

	Studies that use a sink particle algorithm typically use the Jeans density as the main criterion for particle creation. This is the case in, for example, \cite{2009Sci...323..754K}, \cite{2007ApJ...656..959K}, \cite{2016MNRAS.463.2553R}, \cite{2010ApJ...725..134P} and \cite{2011ApJ...729...72P}. As mentioned before, a self-consistent treatment of the fragmentation process, instead of a sink particle algorithm, provides more certainty in the number of fragments formed, and, as we have shown here, in the number of fragments that survive the fragmentation process and can become companions. We have also shown that fragments, while in the hydrostatic core phase, undergo several interactions that can destroy them, and they need to reach a second core status in order to safely replace them with a sink particle. The use of sink particles for hydrostatic cores, additionally misses the formation of the smaller secondary disks, and the fact that their associated spiral arms can occasionally give rise to new fragments.

	The fact that AMR grids with minimum cell sizes of $20\unit{au}$ and $10\unit{au}$ do not resolve adequately the Jeans length, added to the fact that we do not observe fragmentation in runs x1 and x2, suggests that sink particles created in simulations that use such grids, although they represent regions of high density, do not (necessarily) represent physical local collapse. Then, companions produced in this way are due to purely numerical effects, and hence, we do not recommend the use of a sink particle algorithm in multiplicity studies if not used with a grid with adequate resolution.

\subsection{Additional physical effects} \label{S: literature-addphysics}

	Contrary to our setup, the simulations from \cite{2009Sci...323..754K}, \cite{2016MNRAS.463.2553R} (laminar) and \cite{2016ApJ...823...28K} do not restrict the motion of the central massive protostar via a fixed sink cell. However, these studies show qualitatively that if the system forms a disk, the central massive protostar does not excessively move from the center. This was studied in more detail and confirmed in \cite{2019MNRAS.482.5459M}, where they implemented disk wobbling and found no dramatic differences with treating the massive protostar as fixed during the fragmentation period.
	
	With regards to the initial density profile, studies that consider a constant density profile \citep[e.g.][]{2010ApJ...725..134P, 2011ApJ...729...72P, 2012MNRAS.420..613G} show that the cloud forms many disperse fragments, while clouds with power law-density profiles \citep[e.g.][and this work]{2012MNRAS.420..613G, 2016MNRAS.463.2553R, 2018MNRAS.473.3615M} are dominated by a central object, with less fragmentation the steeper the profile is. Observational studies \citep[e.g.][]{2002ApJ...566..945B, 2005A&A...437..947V, 2014ApJ...785...42P} typically find density slopes $\beta_\rho$ of between $-1.5$ and $-2.6$.
	
	From the studies considered in Sect. \ref{S: literature}, \cite{2012MNRAS.420..613G}, \cite{2016MNRAS.463.2553R},  \cite{2011A&A...528A..72H} and \cite{2007ApJ...656..959K} performed simulations with supersonic initial turbulence, while \cite{2009Sci...323..754K}, \cite{2016MNRAS.463.2553R}, \cite{2010ApJ...725..134P}, \cite{2011ApJ...729...72P} and \cite{2016ApJ...823...28K} considered solid body rotation. Furthermore, \cite{2018MNRAS.473.3615M} offered a parameter scan for different initial angular velocity profiles, without turbulence. Highly-supersonic turbulence leads to filamentary accretion that produces fragments with lower mass ratios to the most massive fragment, and are separated by longer distances than in a weakly-turbulent cloud dominated by thermal Jeans fragmentation. The amount of turbulent fragmentation has been found to depend on the level of turbulence, with higher velocity dispersions leading to more fragmentation \citep{2019ApJ...887..108R, 2018A&A...615A..94F}. Also, the most massive protostar moves considerably from the center of the cloud, although a (fragmenting) disk is formed in some cases \citep{2016MNRAS.463.2553R, 2019ApJ...887..108R}. The parameter scan in \cite{2018MNRAS.473.3615M} revealed that a steep initial angular velocity profile, such as the one used here, produces more and earlier fragmentation.

	Our choices for initial conditions are in agreement with the discussion in \cite{2018A&A...617A.100B} and references therein. In that paper, observations of 20 high-mass star-forming regions showed that fragmentation is more consistent with the thermal Jeans fragmentation picture, and so, low values of turbulence are expected.

	Some of the studies mentioned in Sect. \ref{S: literature} consider the presence of magnetic fields. In general, the presence of magnetic fields is expected to reduce fragmentation. In \cite{2018A&A...620A.182K}, the effect of magnetic fields was studied in the context of high resolution, two-dimensional simulations on jet launching. Magnetic pressure was found to dominate the inner parts of the disk, and so, we expect a more stable disk in these regions. This could also mean that fragmentation may be observed later in time, as the disk would need to reach a certain size to become unstable and fragment. Magnetic braking is also expected to reduce the size of the disk, and since we observe fragmentation in the outer disk, this can reduce the number of fragments. For more details on the role of magnetic fields on forming massive stars, see, for example, \cite{2011ApJ...742L...9C}, \cite{2011MNRAS.417.1054S} and \cite{2020AJ....160...78R}.

\subsection{Accretion bursts}

	\cite{2017MNRAS.464L..90M} and \cite{2019MNRAS.482.5459M} describe the existence of accretion bursts, a fact that we have also observed in our simulations, in the same way. Our analysis of the mass of the fragments (see Sect. \ref{S: impl-accb}), their second collapse phase and their possible final orbits leads us to conclude, however, that strong accretion bursts of magnitudes $\gtrsim 4$ as described in \cite{2019MNRAS.482.5459M} are unlikely to happen with the predicted frequency, since the masses needed for these events correspond typically to fragments that should have undergone second collapse and, due to the reduced collisional cross section, are less probable of being accreted directly by the central massive protostar, instead forming companions.

% - - - - - - - - - - - - - - - - - - - - - - - - - - - - - - - - - - - - - - - -
\section{Summary and conclusions} \label{S: summary}

We have presented the highest resolution, three-dimensional self-gravity-radiation-hydrodynamical simulations of a forming and fragmenting accretion disk around a forming massive star performed up to now, incorporating stellar evolution and dust sublimation and evaporation. The high resolutions achieved in this work mean that no sink particles were necessary to describe the fragments, which also has allowed us to study the fragmentation mechanisms and interactions, as well as the structure of fragments. A detailed analysis of the formation, interactions and ultimate fate of the fragments has been presented. For this analysis, sophisticated data post-processing techniques allowed us to track the masses, central temperature of the fragments and orbits. We also studied the further evolution of these fragments into companion stars, by considering the conditions for the formation of second Larson cores.

The same simulation has been run five times with varying resolutions. We have studied convergence in our results, and checked that we resolve properly the disk pressure scale height and the Jeans length.

 In a nutshell, it is found that

\begin{itemize}

\item A growing accretion disk is formed around a central massive protostar. 
\item The disk forms spiral arms, and they in turn fragment.

	\item In the fragmenting disk, the Toomre parameter indicates that the disk is globally stable, while being locally unstable, and therefore an insufficiently resolved disk may appear to be Toomre-stable while fragmentation is happening at lower scales.
	\item Fragments are highly dynamic; they form, interact, can be destroyed, sometimes form secondary disks with their own spiral arms.
	\item Secondary disks can also fragment.
	\item Fragments tend to have highly eccentric and chaotic orbits, due to their continuous interactions. However, they can also gain enough mass to be in more stable orbits and form companions in the middle or outer regions of the disk.
	\item The masses of the fragments are of the order of one solar mass, once their central temperatures can reach the hydrogen dissociation limit ($2000\unit{K}$). Fragments that go beyond that temperature, should form second Larson cores, while others remain as hydrostatic cores.

	\item Fragments in general can migrate inwards, if they are hydrostatic cores, they might be accreted by the central massive protostar and create accretion bursts, and if they become second cores, they would most likely form spectroscopic companions.
	\item The number of fragments that survive the fragmentation epoch is $\lesssim 10$.

	\item Reducing resolution by a factor of two (run x8), does not significantly affect the overall picture of the results, except for the report of lower central temperatures in the fragments.
	\item Reducing resolution by a factor of four (run x4), the formation of secondary spiral arms is limited, and the fragments do not achieve the required temperatures for second collapse any more.
	\item  Reducing the resolution by a factor of eight (run x2), the formation of the primary disk and its spiral arms is recovered, but the fragmentation physics is not properly achieved anymore.
	\item Reducing the resolution by a factor of sixteen (run x1), only the formation and growth of the primary disk is recovered, but its internal substructure cannot be resolved anymore.

\end{itemize}

	% - - - - - - - - - - - - - - - - - - - - - - - - - - - - - - - - - - - - - - - -
	\begin{acknowledgements}
		G.A.O.-M. acknowledges financial support by the Deutscher Akademischer Austauschdienst (DAAD), under the program \textit{Research Grants - Doctoral Programmes in Germany}.

		R.K. acknowledges financial support via the Emmy Noether Research Group on Accretion Flows and Feedback in Realistic Models of Massive Star Formation funded by the German Research Foundation (DFG) under grant No. KU 2849/3-1 and KU 2849/3-2.
		
		We thank the anonymous referee for their very valuable and helpful comments and suggestions.
	\end{acknowledgements}

	\bibliographystyle{aa} % style aa.bst
	\bibliography{fragbib,PapersRolf}

\end{document}